\documentclass[12pt]{article}

\usepackage{epsfig, amssymb}
\usepackage{graphicx,epsfig}
\usepackage{epstopdf}
\usepackage{graphicx}
\usepackage{caption}
\usepackage{subcaption}
\usepackage{subfig}
\usepackage{color}
\usepackage{amsmath}

\begin{document}
\begin{titlepage}
\thispagestyle{empty}
\begin{flushright}
\end{flushright}

\bigskip

\begin{center}
\noindent{\Large \textbf
{Phase transition in anisotropic holographic superfluids with arbitrary $z$ and $\alpha$}}\\ 
\vspace{2cm} \noindent{
Miok Park\footnote{e-mail:miokpark@kias.re.kr}${}^a$, Jiwon Park\footnote{e-mail:minerva1993@gmail.com}${}^b$ and Jae-Hyuk Oh\footnote{e-mail:jack.jaehyuk.oh@gmail.com}}${}^b$

\vspace{1cm}
  {\it
  Korean Institute for Advanced Study, Seoul 02455, Korea${\ }^a$, \\ 
Department of Physics, Hanyang University, Seoul 133-791, Korea${\ }^b$\\
 }
\end{center}

\vspace{0.3cm}
\begin{abstract}
Einstein-dilaton-$U(2)$ gauge field theory is considered in a spacetime characterised by $\alpha$ and $z$, which are the hyperscaling violation factor and the dynamical critical exponent respectively. We obtain the critical values of chemical potential $\mu_c$ that is defined on its boundary dual fluid
and derives phase transition from spatially isotropic to anisotropic phase for the various values of the $\alpha$ and $z$. To do so, we first apply Sturm-Liouville theory and estimate the upper bounds of the critical values of the chemical potential. We also employ a numerical method {in the ranges of $1 \leq z \leq 4$ and $0 \leq \alpha \leq 4$} to check if the Sturm-Liouville method correctly estimates the critical values of the chemical potential. It turns out that the two methods are agreed within 
10 percent error ranges. Finally, we compute free energy {density} of the dual fluid by using its gravity dual and {check} if the system shows phase transition at the critical values of the chemical potential $\mu_c$ for the given parameter region of $\alpha$ and $z$. Interestingly, it is observed that the anisotropic phase is more favoured than the isotropic phase for small values of $z$ and $\alpha$. However, for large values of $z$ and $\alpha$, the anisotropic phase is not favoured.
\end{abstract}
\end{titlepage}

\tableofcontents
\section{Introduction}
AdS/CFT correspondence\cite{Aharony:1999ti} is the greatest discovery last century in string theory and provides a new tool to study various strongly coupled particle field theories. Especially, fluid/gravity duality and AdS/CMT have been applied to many low energy particle theories like conformal/non-conformal fluid dynamics and condensed matter theories. 
Especially obtaining the ratio of the shear viscosity to the entropy density, $\eta/s$ from the holographic model is the most surprising. It turns out that it has the universal value of $\frac{1}{4\pi}$\cite{Policastro:2002se}. 

In the development of AdS/CMT, the observation of several kinds of symmetry breaking mechanism in a gravitational system plays an important role. The first idea that black hole can superconduct was suggested by the observation that RN-AdS black hole is possibly unstable under a complex scalar perturbation below a certain temperature. Below that temperature the gravitational system presents its scalar hair outside of the black hole horizon in \cite{Gubser:2008px}. Based on this mechanism, the holographic superconductor model was established in \cite{Hartnoll:2008kx} which shows a complex scalar field condensation resulting from a spontaneous symmetry breaking of global U(1) and it corresponds to an order parameter in the second phase transition via a holographic interpretation.

Another types of holographic superconductor/superfluidity model was also investigated in the asymptotically charged-AdS$_{4}$ spacetime by employing SU(2) non-Abelian gauge field in \cite{Gubser:2008zu}. Some properties to the holographic dual description such as speed of second sound or the conductivity are studied in asymptotically AdS$_{5}$, by taking the probe limit in \cite{Herzog:2009ci}. This model assumes that a chemical potential is given in the third isospin direction and accordingly it has the response $\langle j^{t}_{3} \rangle$, which breaks global SU(2) symmetry to U(1). Interestingly, below a certain temperature $T_{c}$, additional current starts to be induced in a spatial direction, denoted as $\langle j^{x_{1}}_{1} \rangle$. This current breaks U(1) symmetry and also rotational symmetry of the system to U(1). In the dual field theory it plays a role of the order parameter for the second order phase transition. The holographic dual of the anisotropic fluid dynamics is described by excitations in the background of asymptotically RN-AdS black brane solution obtained from Einstein-$SU(2)$ Yang-Mills theory defined in 5-dimensional space. In the spatially isotropic phase a temporal part of the Yang-Mills fields, $A^3_{t}$ is non-zero only, but in anisotropic phase, a spatial part of the Yang-Mills fields arises, $A^1_{x_1}$ together with the temporal part. In \cite{Policastro:2002se,Herzog:2009ci}, it is found that the the  phase transition occurs at the chemical potential\footnote{This chemical potential $\mu$ is a dimensionless obtained by rescaling with the black brane horizon $r_0$.} $\mu_c=4$ and below the critical temperature the $\langle j^{x_{1}}_{1} \rangle$ starts to appear. Near the critical point where the current $\langle j^{x_{1}}_{1} \rangle$ takes a small value $\epsilon$, the free energy was analytically computed  from the dual gravity side with power expansion of $\epsilon$. It is proved that the anisotropic phase is thermodynamically favourable when $\mu\geq4$.

Beyond considering the asymptotically AdS spacetime, the applications of the holography to strongly correlated systems have inspired to conceive more various gravitational systems such as Lifshitz spacetime, hyperscaling violation geometry and so on. The Lifshitz spacetime is firstly introduced to realise temporal anisotropy emerged from quantum critical phenomena associated with continuous phase transitions in \cite{Kachru:2008yh}. In the vicinity of the critical point, time scales differently from space 
\begin{equation}
t \rightarrow \lambda^{z} t, \qquad \vec{x} \rightarrow \lambda \vec{x}
\end{equation}
where $z$ is the dynamical critical exponent, and its geometrical realisation can be written as
\begin{equation}
ds^{2} = - r^{2z} dt^{2} + \frac{dr^{2}}{r^{2}} + r^{2} d \vec{x}^{2}
\end{equation}
where it restores the conformal invariance when $z=1$. It is known that this geometry can be generated in several ways; considering a massive vector field, adding higher curvature terms or coupling between an Abelian gauge field and a dilaton field into the action. Furthermore, the gravitational action having the Abelian gauge fields coupled to the dilaton field is allowed to make more general extension for the Lifshitz spacetime to have overall hyperscaling factor $\alpha$ and so not to be invariant under scaling
\begin{equation}
ds^{2} \rightarrow \lambda^{- \alpha} ds^{2},
\end{equation}
and the extended metric takes a form of 
\begin{equation}
ds^{2} = r^{2 \alpha} \bigg( - r^{2z} dt^{2} + \frac{dr^{2}}{r^{2}} + r^{2} d \vec{x}^{2} \bigg).
\end{equation}
Some properties of this spacetime are studied in \cite{Alishahiha:2012qu,Charmousis:2010zz}. 

In the previous researches, the  phase transition from the spatially isotropic to anisotropic system is widely studied in the charged-AdS black brane spacetime, which is $z=1$ and $\alpha=0$ case. In this note, we {consider a more general spacetimes having an arbitrary $z$ and $\alpha$ and assume that the system is near the critical temperature $T_{c}$. So the current $\langle j^{x_{1}}_{1} \rangle$ just starts to be induced and take a small value, $\langle j^{x_{1}}_{1} \rangle \sim \epsilon$. Our purpose is to find the critical value of the chemical potential {$\mu$} for generic $z$ and $\alpha$  and to compute free energy density to check the thermodynamically favoured state. To construct such a holographic model, we consider Einstein-dilaton-$U(2)$ theory, which gives the background geometry with asymptotically hyperscaling violation and Lifshitz scale symmetry. Here, for simplicity, we take the limit that the Yang-Mills coupling constant is large and there is no back reaction between the geometry and the Yang-Mills fields. Namely, we consider a probe limit. 

To search the critical values of the chemical potential according to $\alpha$ and $z$, we apply two different methods: analytic and numerical studies. As the analytic approach, we use the Sturm-Liouville theory. The Sturm-Liouville theory is to solve differential equations with some undetermined parameters in the equations. Suppose a differential equation with a parameter $x$. There might be a solution of the equation but the solution exists only when the $x$ becomes an appropriate value. In fact, it is Eigen value problem. One can construct Eigen functions determining the values of the $x$ as their Eigen values and there is an Eigen function which gives the lowest value of the $x$.
This also means that once we suppose there is a solution of the equation, then one can estimate the value $x$ as its upper bound by employing some trial solutions and applying the variational principle\cite{griffith}. For our case, the $x$ corresponds to the $\mu$ and the solution of differential equation does to the spatial part of the Yang-Mills fields. We point out that this method provides the upper bounds of the critical values of the chemical potential.

In order to numerically find a critical value of $\mu$, we solve the coupled Yang-Mills equations of the $A^1_{x_1}$ and $A^3_t$ with appropriate boundary conditions at the black hole horizon and the asymptotics for a fixed value $z$ and $\alpha$ by using a shooting method. Then we compare the Sturm-Liouville results with the numerical one. This is one of our main results, which is given in Fig.\ref{zalphafix}. In Fig.\ref{zalphafix}-(a), we plot the critical values of the chemical potential with the solid lines(analytic approach) and dashed lines(numerics) for $\alpha=0,1,2,3,4$ from below in order. They present monotonically increasing behavior as the $z$ {increases} for the fixed values of $\alpha$. In Fig.\ref{zalphafix}-(b), we plot the critical values of the chemical potential $\mu_c$ as the $\alpha$ increases for values of $z=1,2,3,4,5$ from below in order.

We also derive the free energy from the Euclideanized dual gravity on-shell action and compute it for each value of the $\mu_c$ to compare those of the isotropic state and anisotropic state by using the numerical solutions of Yang-Mills fields. Interestingly, the numerical results show that the anisotropic state is thermodynamically favoured only in the certain area for $\alpha$ and $z$. It turns out that the values of $\alpha$ and $z$ are relatively small in this region. 
 The free energy for the isotropic state is always negative, but for the anisotropic state it is negative only in that small region and slightly larger than the isotropic state there. Apart from the region that has small values for $a$ and $z$, the free energy for the anisotropic state takes positive values and exponentially grows. The detail will be discussed in Sec.\ref{Numerics}. 

This note is organised as follows. In Sec.\ref{Holographic Model}, we discuss our holographic setting of the gravity model which gives asymptotically hyperscaling violation, Lifshitz scaling symmetry and spatial anisotropy for the critical values of the chemical potential. In Sec.\ref{Analytic and numerical approaches to the critical points}, we explain our analytic method and in Sec.\ref{Numerics}, we demonstrate numerical methods and the results. In Sec.\ref{summary}, we summarize our work.

\section{Holographic Model}
\label{Holographic Model}
We start with a bulk action as
\begin{eqnarray}
S&=&\frac{1}{\kappa^2_5}\int d^5 x \sqrt{-g}\left[ R- \frac{1}{2} g^{MN}\partial_M \phi \partial_N \phi
+\frac{V_0}{L^2}e^{\gamma\phi}-\frac{\kappa^2_5}{4g^2_U}e^{\lambda_U\phi}F_{MN}F^{MN}\right. \\ \nonumber
&-&\left.\frac{\kappa^2_5}{4g^2_{YM}}e^{\lambda_{YM}\phi}G^a_{MN}G^{aMN}\right],
\end{eqnarray}
where $M$ and $N$ are 5-dimensional(5-D) spacetime indices, running from 0 to 4,
$g_{MN}$ is spacetime metric, $V_0$, $\gamma$, $\lambda_U$ and $\lambda_{YM}$ are real constants and $\kappa_5$ is
5-D gravity constant. $\phi$ is a real scalar field, and $F_{MN}$ is field strength of U(1) gauge field $A_M$, i.e. $F_{MN}=\partial_M A_N-\partial_N A_M$. $G^a_{MN}$ is field strength of Yang-Mills field $B^a_M$( {$B\equiv B_{M} dx^{M} = \tau^{a} B^{a}_{M} dx^{M}$}), where we have chosen the simplest Yang-Mills gauge group{; SU(2) that satisfies $[\tau^{a}, \tau^{b}] = i \epsilon^{abc} \tau^{c}$ and Tr$(\tau^{a} \tau^{b}) = \delta^{ab}/2$ where $\epsilon$ is fully antisymmetric tensor and $\epsilon^{123}=1$}. 
Then, 
\begin{equation}
G^a_{MN}=\partial_M B^a_N-\partial_N B^a_M-\epsilon^{abc}B^b_MB^c_N,
\end{equation}
where the gauge group adjoint indices, $a,b$ and $c$ run over 1 to 3. $g_U$ and $g_{YM}$ are gauge couplings. We set {the AdS radius to be} $L=1$.

The bulk equations of motion are given by
\begin{eqnarray}
X&\equiv&\frac{1}{\sqrt{-g}}\partial_M(\sqrt{-g}g^{MN}\partial_N\phi)+V_0\gamma e^{\gamma\phi}
-\frac{\kappa^2_5\lambda_U}{4g^2_U}e^{\lambda_U\phi}F_{MN}F^{MN} \\ \nonumber
&-&\frac{\kappa^2_5\lambda_{YM}}{4g^2_{YM}}e^{\lambda_{YM}\phi}G^a_{MN}G^{aMN}=0, \\
Y^N&\equiv&\frac{1}{\sqrt{-g}}\partial_M(\sqrt{-g}e^{\lambda_U\phi}F^{MN})=0, \\
\mathcal{Y}^{aN} &\equiv&\frac{1}{\sqrt{-g}}\partial_M(\sqrt{-g}e^{\lambda_{YM}\phi}G^{aMN})+e^{\lambda_{YM}\phi}\epsilon^{abc}G^{bMN}B^{c}_M=0, \\ 
W_{MN}&\equiv&R_{MN}+\frac{V_0}{3}g_{MN}e^{\gamma\phi}-\frac{1}{2}\partial_M \phi \partial_N \phi-\frac{\kappa^2_5}{2g^2_U}e^{\lambda_U\phi}\left( F_{PM}F^P_N
-\frac{1}{6}g_{MN}F_{PQ}F^{PQ} \right) \\ \nonumber
&-&\frac{\kappa^2_5}{2g^2_{YM}}e^{\lambda_{YM}\phi}\left( G^a_{PM}G^{aP}_{\ \ N}
-\frac{1}{6}g_{MN}G^a_{PQ}G^{aPQ} \right)=0.
\end{eqnarray}
We start with a solution having generic hyperscaling violating factor $\alpha$ and temporal anisotropy factor $z$. Such solutions already appeared in \cite{Alishahiha:2012qu}, an Einstein-dilaton theory with two different $U(1)$ gauge fields. This is spatially isotropic solution. Our solution is obtained by the similar footing but we want to get spatial anisotropy on top of this. The ansatz is given by
\begin{eqnarray}
\label{ansatz}
ds^2&=&r^{2\alpha}\left( -r^{2z}f(r)\sigma^2(r)dt^2 + \frac{dr^2}{r^2f(r)} +r^2h^{-4}(r)dx^2_1
+r^2h^2(r) (dx^2_2+dx^2_3)\right) \\ \nonumber
B^a\tau^a&=&b(r)\tau^3dt+\omega(r)\tau^1dx_1, \\ \nonumber
A&=&a(r)dt \\ \nonumber
\phi&=&\phi(r),
\end{eqnarray}
where $\omega(r)$ represents the spatial anisotropy and $\tau^a=\frac{\sigma^a}{2}$. When there is no anisotropy, i.e. $\omega(r)=0$, the background geometry becomes 
5-D black brane solutions which are given by
\begin{eqnarray}
f(r)&=&1-mr^{-3\alpha-z-3}+\frac{\kappa^2_5}{g^2_{YM}}\tilde\mu^2r^{-6\alpha-4-2z}, {\ \ }\sigma(r)=1, {\ \ }h(r)=1 \\ \nonumber
G^3_{rt}&=&\partial_r b(r)=\tilde\mu e^{-\sqrt{\frac{\alpha+z-1}{6(\alpha+1)}}\phi_0}\sqrt{6(\alpha+1)(3\alpha+z+1)}r^{-3\alpha-z-2}, {\ \ }G^1_{rx_1}=0, \\ \nonumber
F_{rt}&=&\partial_r a(r)=\frac{g^2_U}{\kappa^2_5}e^{\frac{2\alpha+3}{\sqrt{6(\alpha+1)(\alpha+z-1)}}\phi_0}\sqrt{2(z-1)(3\alpha+z+3)}r^{3\alpha+z+2}, \\ \nonumber
\phi&=&\phi_0+\sqrt{6(\alpha+1)(\alpha+z-1)}\ln r,
\end{eqnarray}
where $m$ is mass density of the black brane and $\tilde\mu$ is chemical potential, and the free parameters of this model are fixed by 
\begin{equation}
{V_{0} = (3 \alpha +z+2) (3 \alpha +z+3) e^{\gamma \phi_{0}}}
\end{equation}
and 
\begin{equation}
\gamma=-\frac{{\sqrt{2}}\alpha}{\sqrt{{3}(\alpha+1)(\alpha+z-1)}},\; \; \lambda_{YM}=\sqrt{\frac{2(\alpha+z-1)}{3(\alpha+1)}}, \; \; \lambda_U=-\frac{2(2\alpha+3)}{\sqrt{6(\alpha+1)(\alpha+z-1)}}.
\end{equation}
 We note that this solution should satisfy null-energy condition\cite{Alishahiha:2012qu}, 
\begin{equation}
\label{null-condition}
(\alpha+1)(\alpha+z-1)\geq0.
\end{equation}
To be more precise, we consider a null vector in this black brane background as $\zeta^M=(\sqrt{g^{rr}},\sqrt{g^{tt}},0,0,0)$, then
\begin{equation}
T_{MN}\zeta^M\zeta^N\sim R^r_r-R^t_t=3(\alpha+1)(\alpha+z-1)r^{-2\alpha}f(r)\geq0
\end{equation}
and this leads (\ref{null-condition}).

We would like to explore a spatial anisotopy in this background by turnning on ${B}^1_{x_{1}}$($=\omega(r)$) without considering its backreactions to the background geometry\footnote{We will leave this project for our future work}. This limit can be obtained by demanding that Yang-Mill's coupling is taken to be infinity, i.e.  $\frac{\kappa^2_5}{g^2_{YM}}\rightarrow 0$.
In such limit, spacetime becomes just AdS-{Swarzschild} type black brane since the term being proportional to chemical potential in $f(r)$ disappears.

For further discussion, we rescale the radial coordinate $r$ by the size of black brane horizon.
 More precisely, we define a new raidal variable $u$ as
\begin{equation}
r=r_0 u,
\end{equation}
where $r_0$ is horizon, which is obtained by 
\begin{equation}
f(r=r_0)=1-\frac{m}{r^{3\alpha+z+3}_0}=0,
\end{equation}
then $m=r^{3\alpha+z+3}_0$. Together with this, we rescale the other coordinate variables as
$t\rightarrow r^{-z}_0t$ and $\{x_1,x_2,x_3\}\rightarrow \frac{1}{r_0}\{x_1,x_2,x_3\}$. As a result, the background metric becomes
\begin{equation}
\label{mtrcprb1}
\frac{ds^2}{r^{2\alpha}_0}=u^{2\alpha}\left( -u^{2z}f(u)dt^2 + \frac{du^2}{u^2f(u)} +u^2(dx^2_1
+dx^2_2+dx^2_3)\right), 
\end{equation}
where
\begin{equation}
\label{mtrcprb2}
f(u)=1-u^{-3\alpha-z-3}.
\end{equation}
Together with this, we define a new chemical potential, $\mu$, in this rescaled coordinate as
\begin{equation}
\mu\equiv\frac{\tilde \mu}{r^{3\alpha+z+1}_0} e^{-\sqrt{\frac{\alpha+z-1}{6(\alpha+1)}}\phi_0}\frac{\sqrt{6(\alpha+1)(3\alpha+z+1)}}{3\alpha+z+1}.
\end{equation}
Then, the bakcground value of ${B}^3_t$ becomes
\begin{equation}
\label{zeroth-b-sol}
b(u)=\mu(1-u^{-3\alpha-z-1}).
\end{equation}
In this rescaled coordinate, Yang-Mill's equations are written as
\begin{eqnarray}
\label{y1-eq}
\mathcal Y^1_x&=&u^{-5\alpha-z-2}\partial_u(u^{3\alpha+3z}f(u)\partial_u\omega(u))+f^{-1}(u)u^{-2\alpha-4}b^2(u)\omega(u)=0 \\ 
\label{y2-eq}
{\rm \ and \ }\mathcal Y^3_t&=&u^{-5\alpha-z-2}\partial_u(u^{3\alpha+z+2}\partial_u b(u))-f^{-1}(u)u^{-2\alpha-4}\omega^2(u)b(u)=0.
\end{eqnarray}

\section{Analytic and numerical approaches to the critical points}
\label{Analytic and numerical approaches to the critical points}
In this subsection, we estimate the critical value of chemical potential for the generic $z$ and $\alpha$.
It is more convenient to use $\xi$ coordinate, where $\xi$ is given by $\xi=\frac{1}{u}$. In this coordinate, (\ref{y1-eq}) and (\ref{y2-eq})
are given by 
\begin{eqnarray}
\label{omega-eq-in-xi}
0&=&\frac{d}{d\xi}\left( \xi^{2-3\alpha-3z}f(\xi)\frac{d\omega(\xi)}{d\xi}\right)
+f^{-1}(\xi)\xi^{-3\alpha-z}b^2(\xi)\omega(\xi), \\ 
\label{b-eq-in-xi}
0&=&\frac{d}{d\xi}\left( \xi^{-3\alpha-z}\frac{db(\xi)}{d\xi}\right)
-f^{-1}(\xi)\xi^{-3\alpha-z}\omega^2(\xi)b(\xi).
\end{eqnarray}
Near critical point, we expect second order phase transition and can regard that the solution of $b(\xi)$ is approximately described by (\ref{zeroth-b-sol}) in the new coordinate
$\xi$ as
\begin{equation}
\label{asymptotics-b}
b(\xi) \sim \mu(1-\xi^{3\alpha+z+1}).
\end{equation}
Near the boundary of the spacetime, $\omega(\xi)$ will behave as
\begin{equation}
\label{asymptotics-omega}
\omega(\xi)=\langle J^1_{x_{1}} \rangle \xi^{3\alpha+3z-1} F(\xi),
\end{equation}
where $\langle J^1_{x_{1}} \rangle$ is an anisotropic order parameter and $F(\xi)$ is a new function that satisfies the following boundary conditions
\begin{equation}
F(0)=1, {\rm\ \ and \ \ } F^\prime(0)=0. \label{eq:Fbc}
\end{equation}

To estimate the critical values of chemical potential for various $\alpha$ and $z$, we use Sturm-Liouville technique. To solve the Sturm-Liouville problem, (\ref{omega-eq-in-xi}) should be written in the form of
\begin{equation}
0=\frac{d}{d\xi}\left( K(\xi)\frac{dF(\xi)}{d\xi} \right)-P(\xi)F(\xi)+\mu^2Q(\xi)F(\xi),
\end{equation}
where
\begin{eqnarray}
K(\xi)&=&\xi^{3\alpha+3z}(1-\xi^{3\alpha+z+3}), \\ 
P(\xi)&=&-(3\alpha+3z-1)\xi^{3\alpha+3z-1}\frac{d}{d\xi}\left( 1-\xi^{3\alpha+z+3} \right), \\
Q(\xi)&=&\frac{\xi^{3\alpha+5z-2}(1-\xi^{3\alpha+z+1})^2}{1-\xi^{3\alpha+z+3}}.
\end{eqnarray}
One can estimate the chemical potential by obtaining its upper bounds by using the general variation method of Sturm-Liouville problem.
In the range of $0\leq \xi \leq 1$, the eigenvalue $\mu^2$ is minimized by the following expression:
\begin{equation}
\mu^2=\frac{\int^1_0 d\xi\left[ K(\xi)\left( \frac{dF(\xi)}{d\xi} \right)^2+P(\xi)F^2(\xi)  \right]}{\int^1_0 d\xi Q(\xi)F^2(\xi)}.
\end{equation}
To estimate this effectively, we introduce a test function of $F(\xi)$ {satisfying \eqref{eq:Fbc}} as
\begin{equation}
F(\xi)=1-t\xi^2,
\label{testftn}
\end{equation} 
where $t$ is an arbitrary real constant to be determined under the condition that $\mu^2$ becomes minimum. The integrations can be analytically performed in the range of
\begin{equation}
5z+3\alpha > 1,{\ \ } 3\alpha+z > -3,{\ \ }4z+6\alpha > -2 {\rm \ \ and\ \ }  z+\alpha > -1
\end{equation}
and on top of this we also consider the null energy condition{(\ref{null-condition})}. Only in this region, the Sturm-Liouville problem is well defined. The range of $z$ and $\alpha$ that we can study is addressed in Figure.\ref{range}, which are regions 1 and 2.

\begin{figure}[h!]
\centering
\includegraphics[width=120mm]{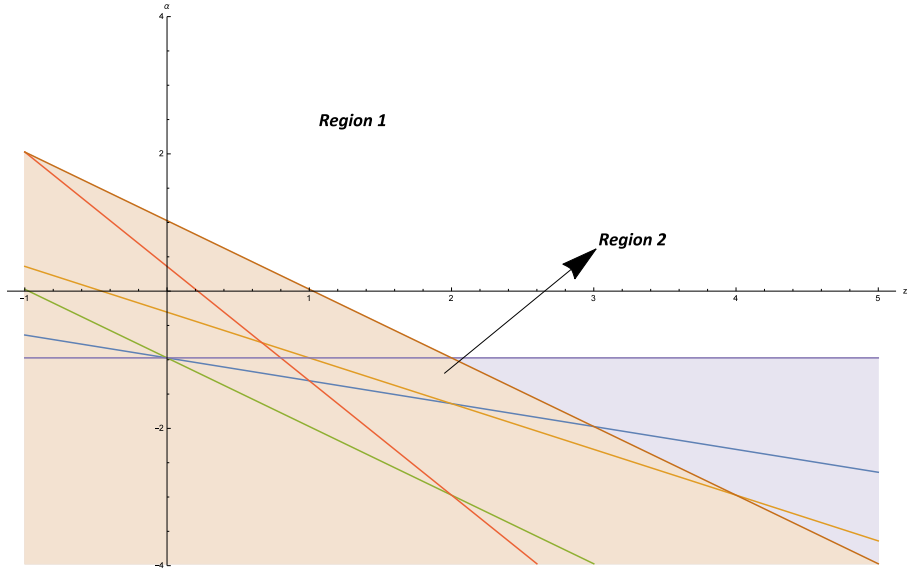}
\caption{Region1 and Region2 are the range of $z$ and $\alpha$ that we apply Sturm-Liouville method.}
\label{range}
\end{figure}

We study the minimum value of the chemical potential for various possible values of $z$ and $\alpha$. 
First of all, we evaluate the upper bounds of the critical value of the chemical potential for fixed $\alpha$. The results are addressed in Fig.\ref{zalphafix}-(a) with solid lines. There are 5 different graphs in it and from below, each solid line indicates the critical value of the chemical potential when $\alpha=0$, $\alpha=1$...$\alpha=4$ as $z$ continuously varies from {1} to {5}. The graphs show monotonically increasing behaviors as
z increases.

Especially, when $z=1$ and $\alpha=0$, we get $\mu_c=4.09206$ by the {Sturm}-Liouville method. In \cite{Basu:2011tt}, the authors address that Einstein-$SU(2)$ Yang-Mills system in asymptotically AdS space has phase transition
from spatially isotropic phase to anisotrophic phase when $\mu_c=4$. The critical value of the chemical potential is correct within 2.3$\%$ error.

Next, we study on the chemical potential upper bounds for fixed $z$ cases. The result is presented in Fig.\ref{zalphafix}-(b) with solid lines too. For large $z$, $\mu_c$ seems to increase linearly as z increases, but for small $z$ region each graph may show minimum of the critical value of chemical potential and rebounds as $\alpha$ decreases.

\begin{figure}[h!]
	\begin{center}
		\begin{subfigure}[b]{0.45\textwidth}
		\includegraphics[scale=0.6]{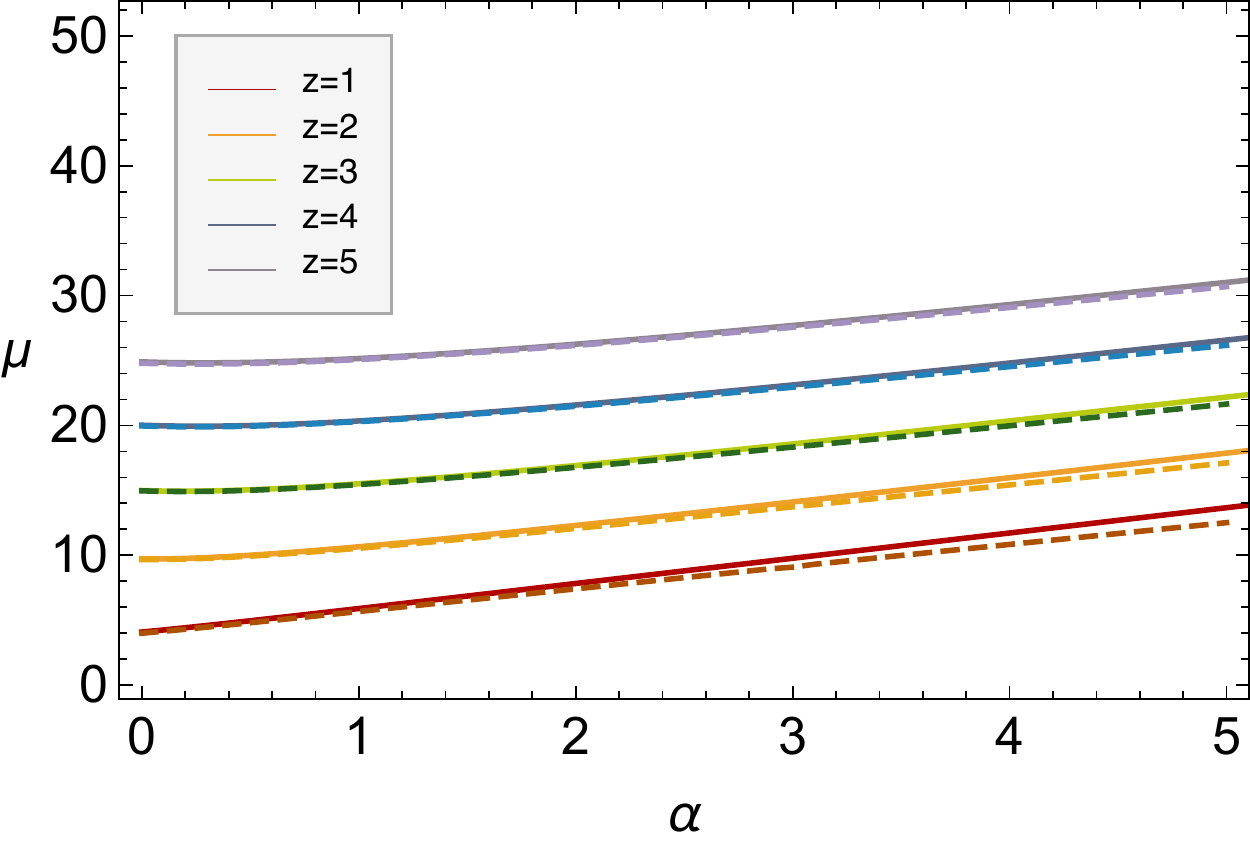}
		\caption{For fixed $z$}
	\end{subfigure} $\; \; \; $
	\begin{subfigure}[b]{0.45\textwidth}
		\includegraphics[scale=0.6]{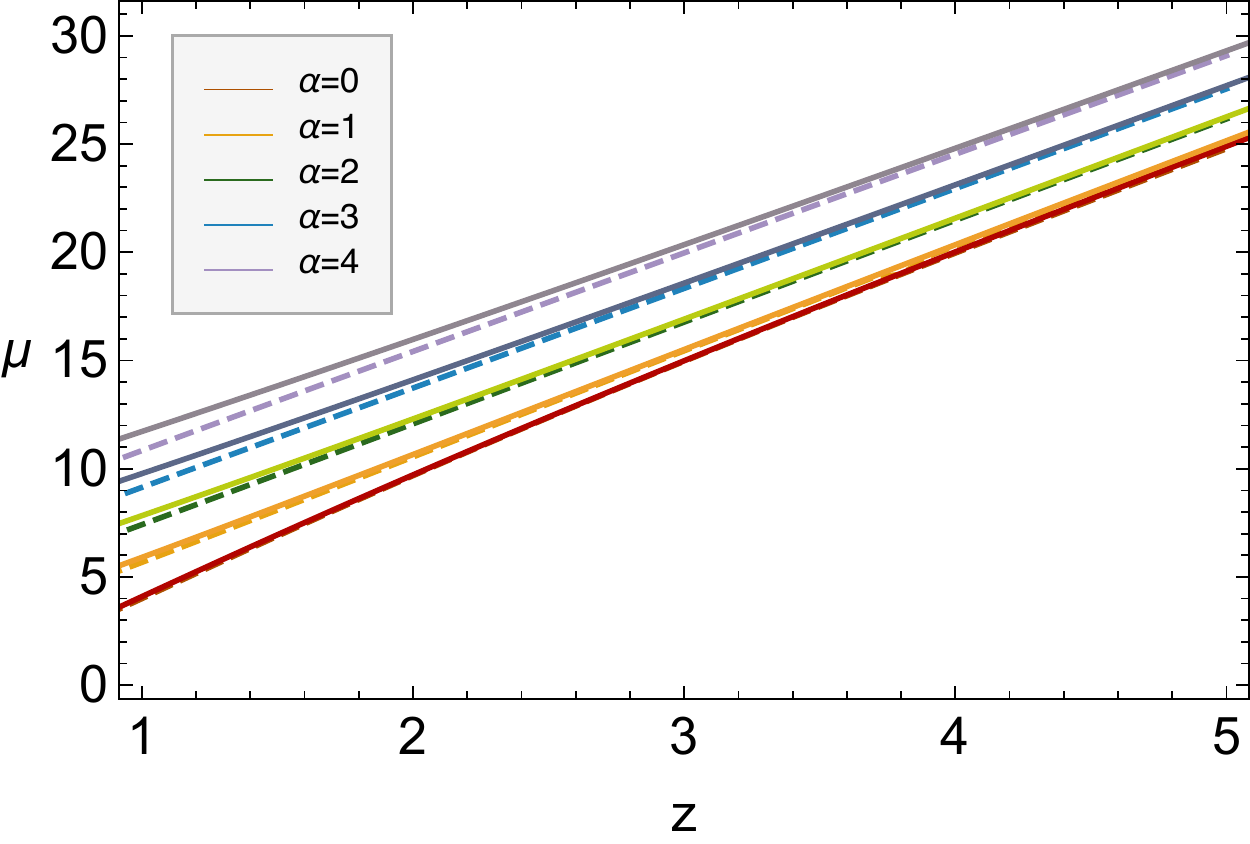}
		\caption{For fixed $\alpha$}
	\end{subfigure}
	\end{center}
\caption{Plot for the chemical potential $\mu$ versus $\alpha$ (or $z$) for the given values of $z$ (or $\alpha$). The solid lines show graphs of the upper bounds obtained from the analytic methods and the dashed lines present the numerical results }
\label{zalphafix}
\end{figure}

\section{Numerics}
\label{Numerics}
In this section, we also consider the limit that the Yang-Mills coupling constant is large, $\frac{\kappa^2_5}{g^2_{YM}}\rightarrow0$, and there is no back reaction to the background geometry, the scalar field and $U(1)$ gauge fields, namely we take the probe limit. 
We numerically find critical values of $\mu$ according to $\alpha$ with fixed $z$ or according to $z$ with fixed $\alpha$, and compare the results with the analytic ones obtained in Sec.\ref{Analytic and numerical approaches to the critical points}. 
We calculate and compare free energy densities with/without turning on $\omega(\xi)$  by using the numerically obtained values of $\mu$, $\alpha$, and $z$. 

\subsection{Numerical solutions of $\omega(\xi)$ and $b(\xi)$}

\paragraph{Initial conditions}
The near horizon expansions on $b(\xi)$ and $\omega(\xi)$ functions satisfying (\ref{omega-eq-in-xi}) and (\ref{b-eq-in-xi}) are given by
\begin{align}
&b(\xi) \sim \mu \left(1-\xi ^{3 \alpha +z+1}\right) + \frac{\mu \left(a_0^2+(3 \alpha +z+3) (3 \alpha +z)^2\right) \left(1-\xi ^{3 \alpha +z+1}\right)^2}{2 (3 \alpha +z+1) (3 \alpha +z+3)} + \mathcal{O}({\varepsilon}^{3}), \\
&\omega(\xi) \sim a_0 -\frac{a_0 \mu^2 \left(1-\xi ^{3 \alpha +z+1}\right)^2}{4 (3 \alpha +z+3)^2} + \mathcal{O}({\varepsilon}^{3}),
\end{align}
where we take the expansion parameter to be $1-\xi^{3 \alpha + z+1}\equiv\varepsilon \ll 1$ because the $\varepsilon$ is the good near horizon expansion parameter when $\omega(\xi)=0$ so we utilize the same one even for $\omega(\xi)\neq0$ case.
The near horizon value of $b(\xi)$(as $\xi \rightarrow 1$) is determined by $\mu, \alpha, z$ and $a_{0}$ and $a_{0} = \omega(1)$. 
Because we assume that the magnitude of the field $\omega(\xi)$ is small(in the language of the dual field theory, the anisotropic order parameter $\langle j^{x_{1}}_{1} \rangle$ is small), we choose $a_{0} = 10^{-5}$ for numerical computation. 
In the asymptotic region(as $\xi \rightarrow 0$), the following boundary conditions are applied:
\begin{align}
\lim_{\xi \rightarrow 0} b(\xi) \sim \mu, \qquad \lim_{\xi \rightarrow 0} \omega(\xi) \sim 0.
\end{align}
We employ shooting method to find the values of $\mu$ and $z$ (or $\mu$ and $\alpha$) for a given value of $\alpha$ (or $z$)
with these boundary conditions.

\paragraph{The numerical results and comparison with the Sturm-Liouville computations} 

The numerical results are presented by the dashed line in Fig.\ref{zalphafix}.
In Fig.\ref{zalphafix}, $\mu$ versus $\alpha$ and $\mu$ versus $z$ graphs are depicted in (a) and (b) for the given integral values of $z=1,2,3,4,5$ and $\alpha = 0,1,2,3,4$  respectively. The solid lines present the results from the Sturm-Liouville method and we show the dashed and solid lines together for comparison. For the given testing range of $\alpha$ and $z$, the discrepancy ratio\footnote{We define the discrepancy ratio $= \frac{\mu_{\rm numeric}-\mu_{\rm Strum-Liuville}}{\mu_{\rm numeric}}\times {\rm 100 \%}$}
between the numerical method and the Sturm-Liouville method 
for $\mu$  is less then {9.2} percents for $0<\alpha<{4}$ and $z={\rm1,2,3,4\ and\ 5}$. It is less then {8.1} percents $\alpha={\rm1,2,3,4\ and\ 5}$ and ${1}<z<10$. 
Especially, for $z=1$ and $\alpha=0$ our numerical value of $\mu_{c}$ yields {3.999}, which well agrees with the result in \cite{Policastro:2002se,Herzog:2009ci}.
As seen in Fig.\ref{fig:dffmu}, the Sturm-Liouville results approach the numerical ones as $z$ becomes larger and $\alpha$ becomes smaller. 

Considering more appropriate test functions may {enhances} the accuracy of the Sturm-Liouville method.
For example, one can go further by adding more terms like having higher power of $\xi$ and find more accurate results.

\begin{figure}[h!]
	\begin{center}
		\begin{subfigure}[b]{0.45\textwidth}
		\includegraphics[scale=0.61]{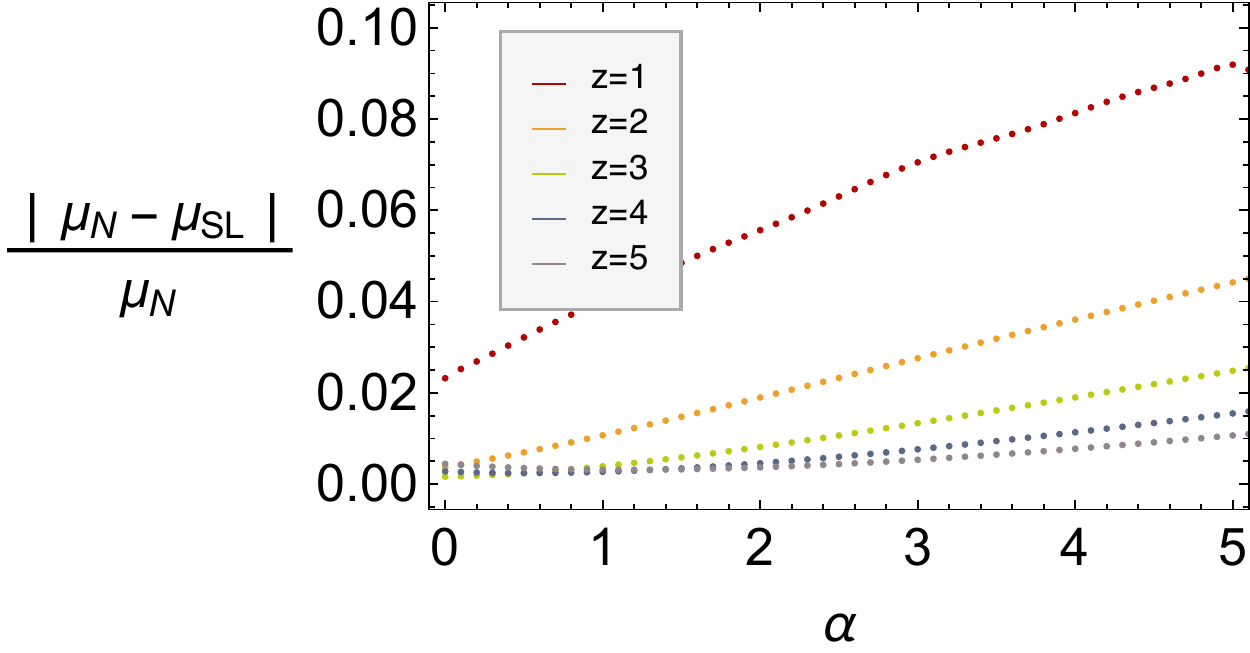}
		\caption{For fixed $z$}
	\end{subfigure} $\; \; \; $
	\begin{subfigure}[b]{0.45\textwidth}
		\includegraphics[scale=0.61]{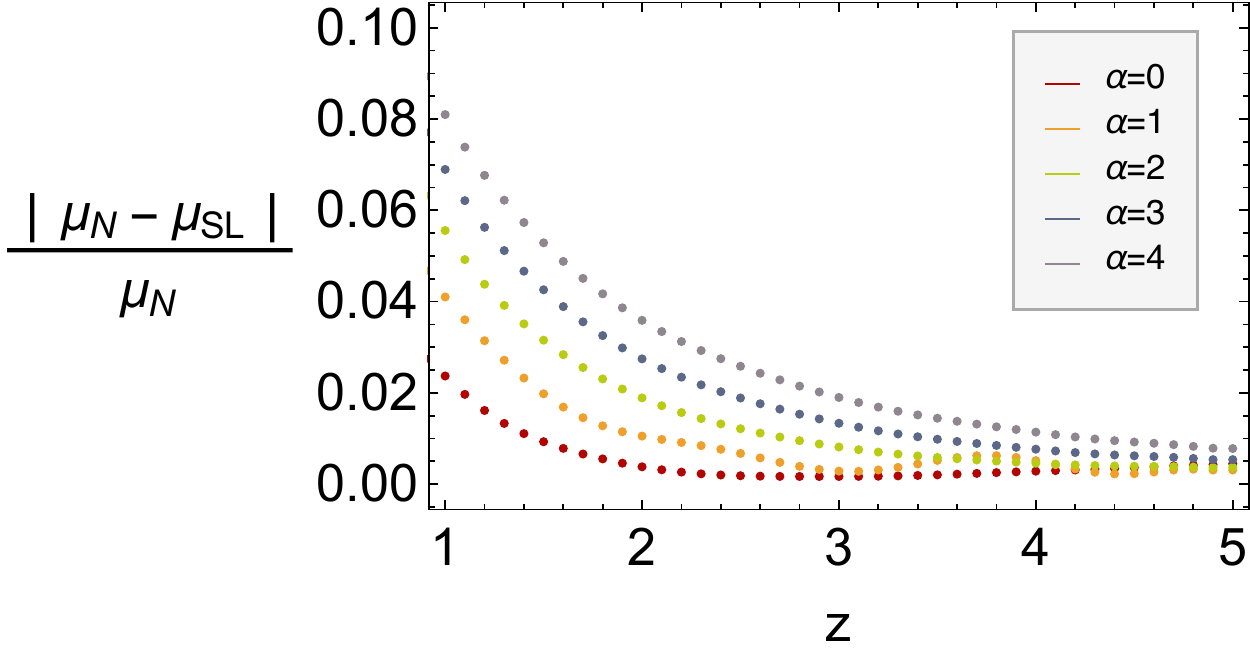}
		\caption{For fixed $\alpha$}
	\end{subfigure}
	\end{center}
\caption{A comparison of the values of $\mu$ in the Sturm-Liouville method and the numerical results.  }
\label{fig:dffmu}
\end{figure}

\begin{figure}[h!]
	\begin{subfigure}[b]{0.32\textwidth}
		\includegraphics[scale=0.44]{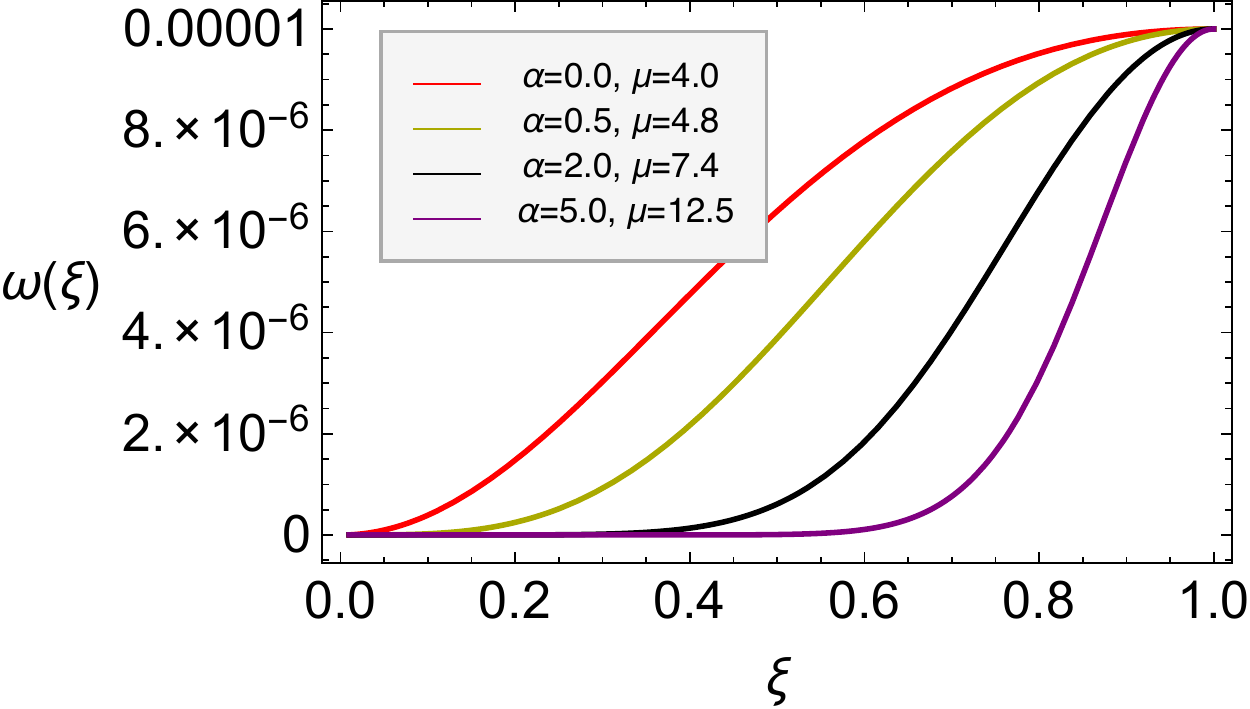}
		\caption{For $z=1$}
	\end{subfigure}	
	\begin{subfigure}[b]{0.32\textwidth}
		\includegraphics[scale=0.44]{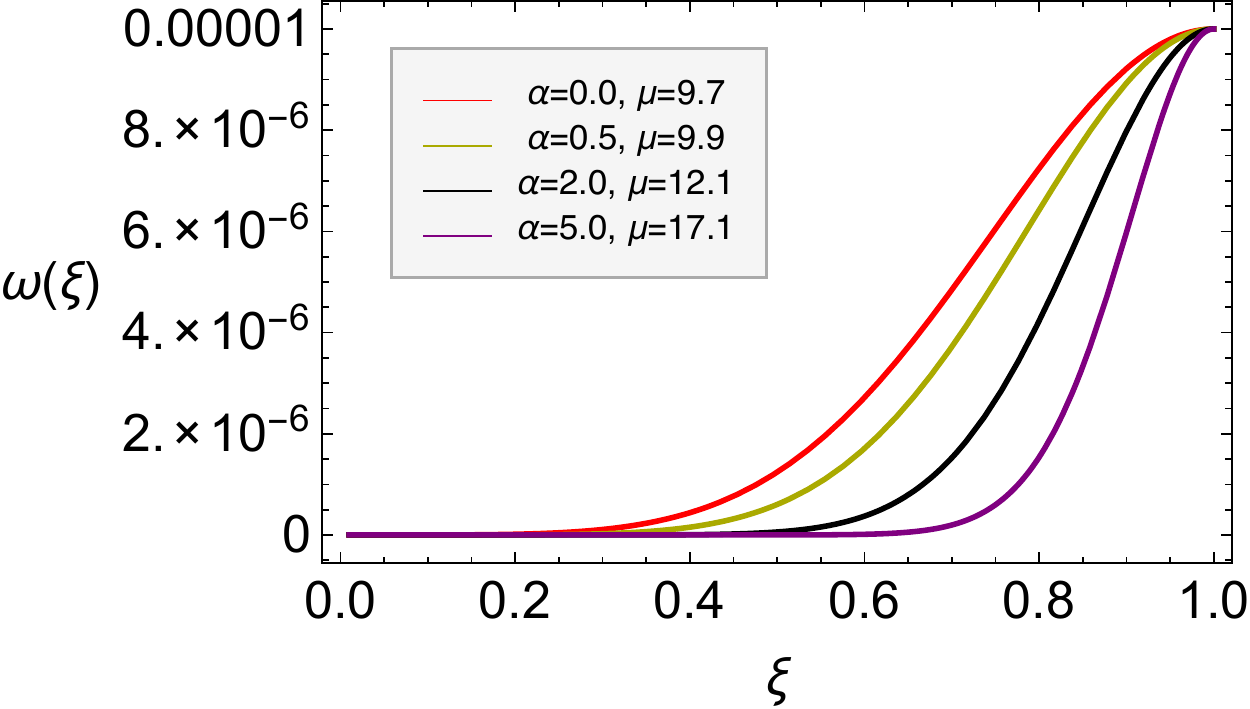}
		\caption{For $z=2$}
	\end{subfigure} 
	\begin{subfigure}[b]{0.32\textwidth}
		\includegraphics[scale=0.44]{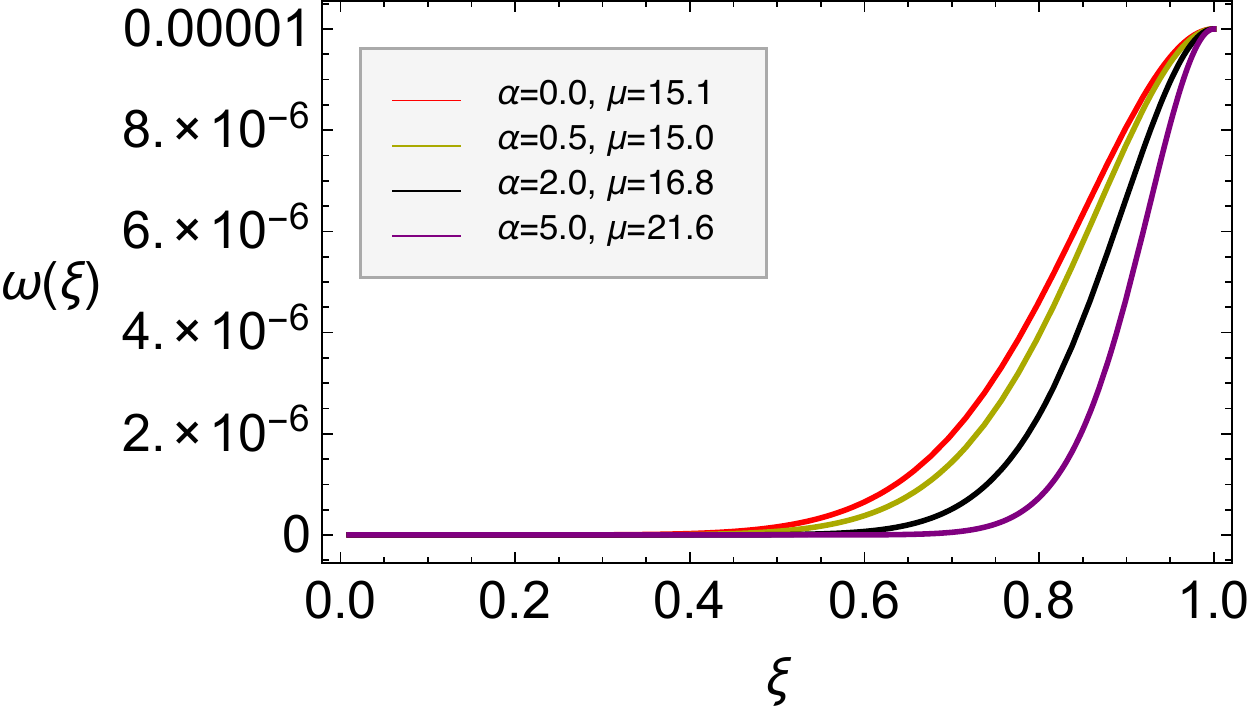}
		\caption{For $z=3$}
	\end{subfigure}\\
	$\;$\\
	\begin{subfigure}[b]{0.32\textwidth}
		\includegraphics[scale=0.4]{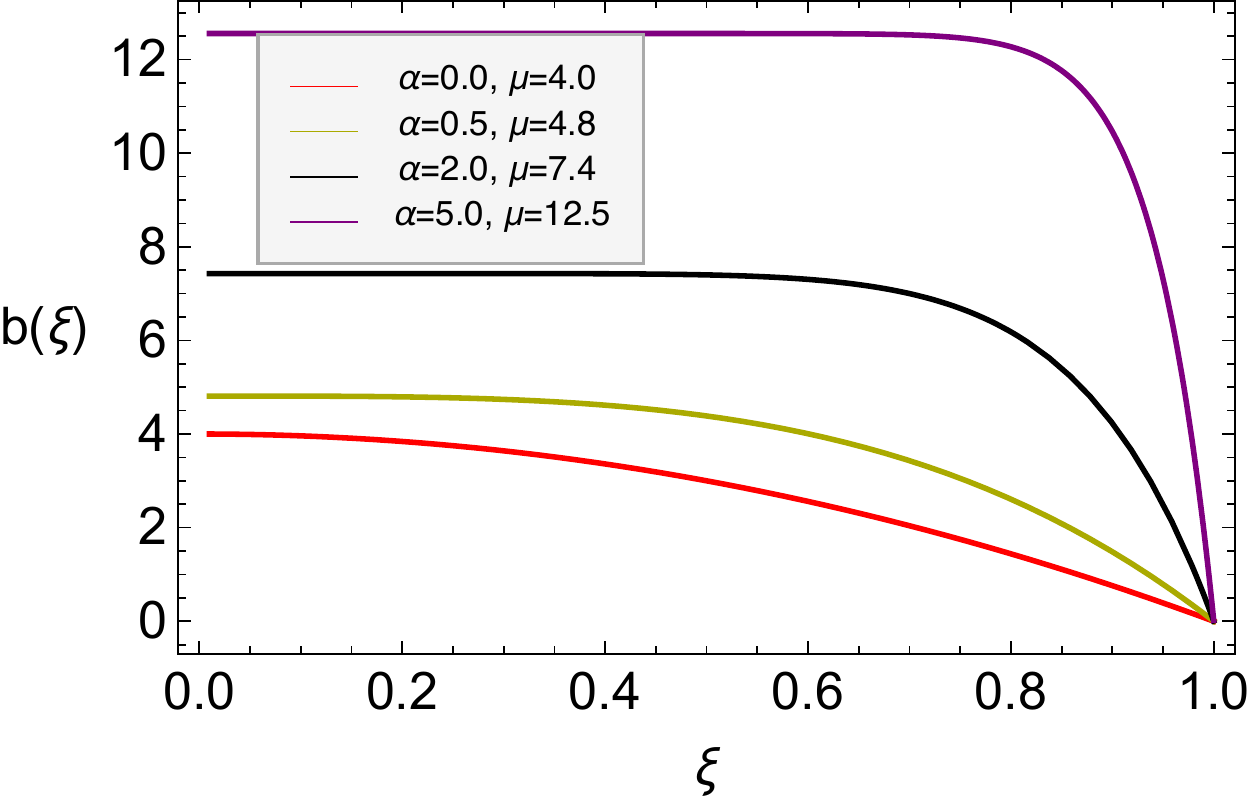}
		\caption{For $z=1$}
	\end{subfigure}	
	\begin{subfigure}[b]{0.32\textwidth}
		\includegraphics[scale=0.4]{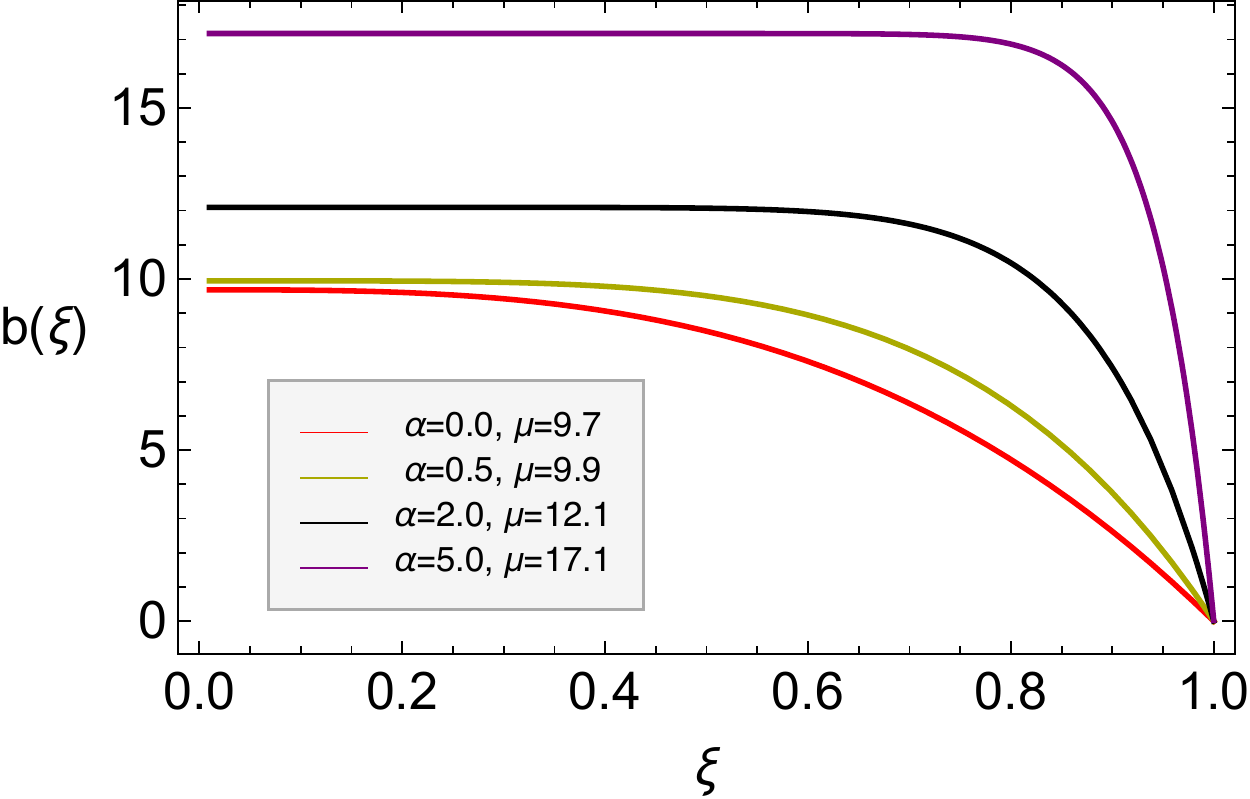}
		\caption{For $z=2$}
	\end{subfigure} 
	\begin{subfigure}[b]{0.32\textwidth}
		\includegraphics[scale=0.4]{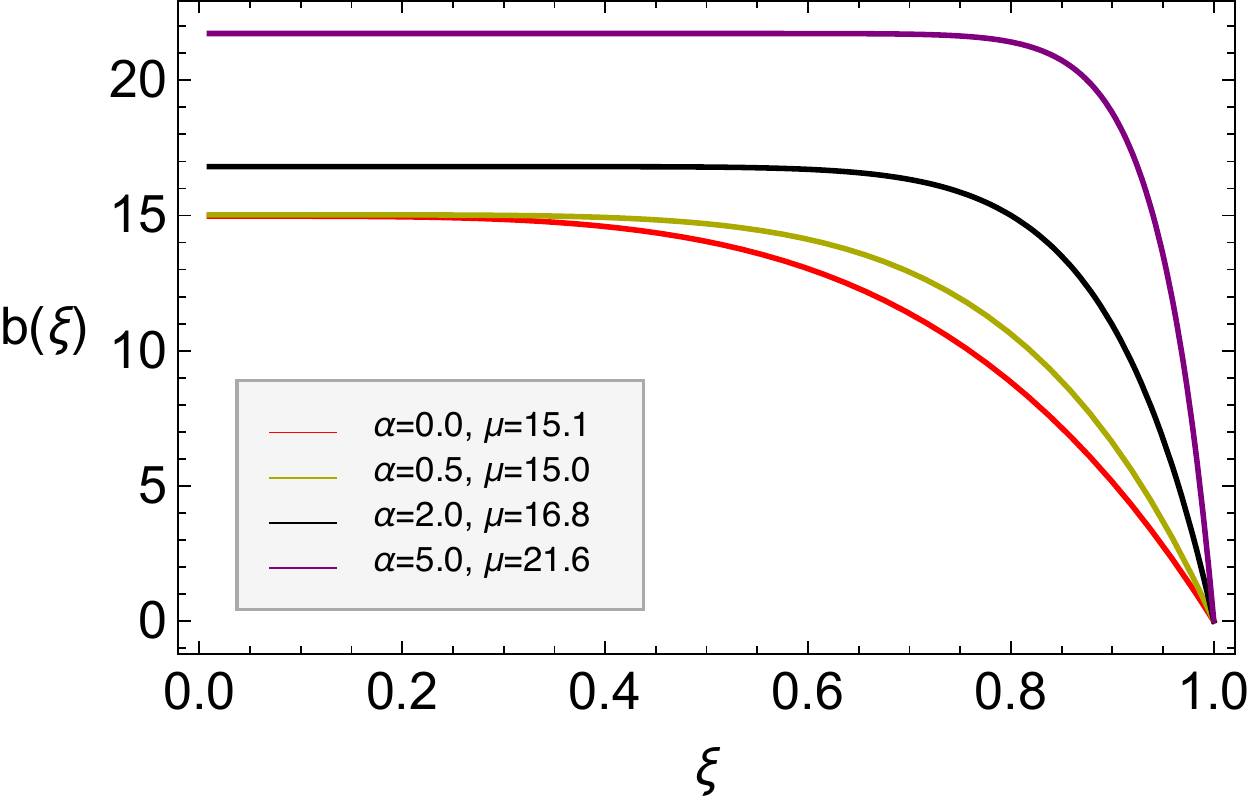}
		\caption{For $z=3$}
	\end{subfigure}
\caption{Numerical fittings of $\omega(\xi)$ and $b(\xi)$ functions.}
\label{fig:wbftnsa1}
\end{figure}

\paragraph{The numerical solutions of the $\omega(\xi)$ and $b(\xi)$ fields} The numerical solutions on $\omega(\xi)$ and $b(\xi)$ are demonstrated in Fig.\ref{fig:wbftnsa1}. 

\subsection{Free energy density}

Since we consider the probe limit, it is enough to inspect the Yang-Mills kinetic term in the background of other fields to examine the thermodynamic phase transition. The Yang-Mills action with imaginary time obtained by Wick rotation as $t = -i\tau$ is given by
\begin{align}
S^E_{YM}&=\frac{\beta V_3}{2g^2_{YM}}r^{3\alpha+2z-2}_0e^{\lambda_{YM}\phi_0}\int^1_0d\xi \xi^{-3\alpha-z} \left[\xi^{2-2z}f(\xi)\left(\frac{d\omega(\xi)}{d\xi}\right)^2
-\frac{\omega^2(\xi)b^2(\xi)}{f(\xi)}-\left(\frac{db(\xi)}{d\xi}\right)^2 \right] \nonumber\\
&\equiv\frac{1}{g^2_{YM}}r^{3\alpha+2z-2}_0e^{\lambda_{YM}\phi_0}
\mathcal S^E_{YM} 
\end{align}
where $V_3$ is the coordinate volume of the spatial boundary. $\beta$ is the periodicity of the Euclidean time in the rescaled coordinate($\xi$ coordinate) and is the inverse of the temperature. By using the equations of motion of the Yang-Mills fields, one can derive simpler form of the Yang-Mills action. The free energy is defined by the Euclideanized on-shell action times the temperature, which is given by
\begin{equation}
\label{FEnergy}
F_{YM} = T \mathcal S^E_{YM}= \frac{V_3}{2}\left(\left.\int^1_0d\xi \xi^{-3\alpha-3z+2}f(\xi)(\partial_\xi\omega(\xi))^2+\xi^{-3\alpha-z}b(\xi)\partial_\xi b(\xi)\right|^{\xi=0}\right).
\end{equation}
In the spatially isotropic phase, $\omega(\xi)=0$, $b(\xi)=\mu(1-\xi^{3\alpha+z+1})$ and so the free energy is given by
\begin{eqnarray}
F_{YM-iso} = T \mathcal S^E_{YM-iso}&=& - \mu^2  \frac{V_3}{2}\int^1_0d\xi(3\alpha+z+1)^2\xi^{3\alpha+z}, \\ \nonumber
&=& - \mu^2  \frac{V_3}{2}(3\alpha+z+1)(1-\xi^{3\alpha+z+1})|^{\xi=0}.
\end{eqnarray}
If we restrict our study within a region, $3\alpha+z+1>0$, then
\begin{equation}
F_{YM-iso}= - \mu^2 \frac{V_3}{2}(3\alpha+z+1).
\end{equation}
If $\Delta F\equiv T \Delta S^E= T  S^E_{YM} - T  S^E_{YM-iso} <0 $, the spatially anisotropic phase is more favoured and then there will be a thermodynamic phase transition from the spatially isotropic phase to the anisotropic one.

\begin{figure}[!b]
	\begin{center}
	\begin{subfigure}[b]{0.45\textwidth}
		\includegraphics[scale=0.64]{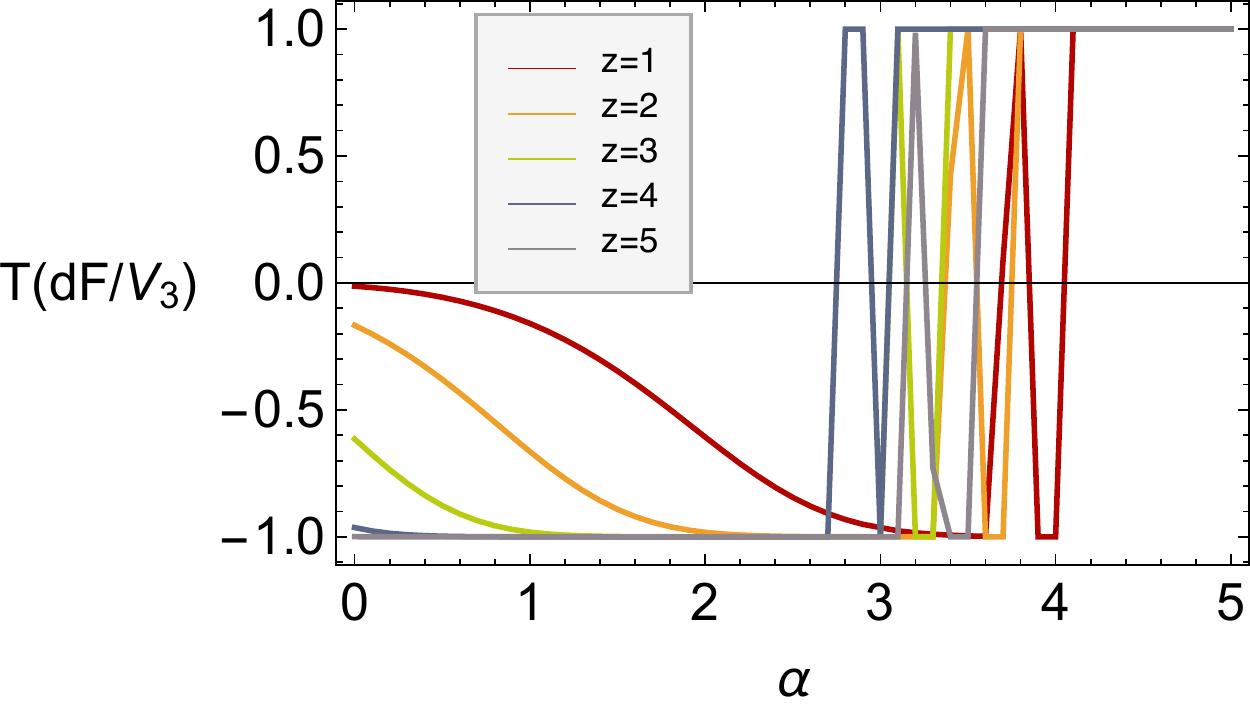}
		\caption{For fixed $z$}
	\end{subfigure} $\; \; \; $
	\begin{subfigure}[b]{0.46\textwidth}
		\includegraphics[scale=0.64]{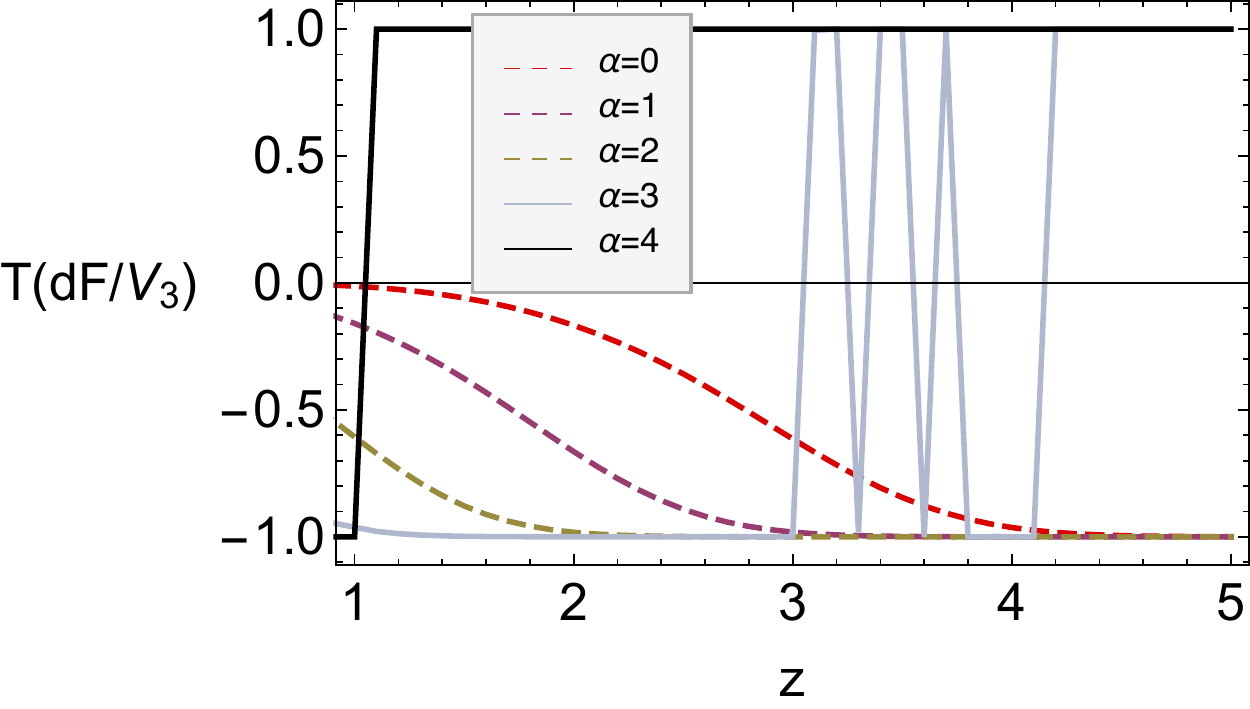}
		\caption{For fixed $\alpha$}
	\end{subfigure}
	\caption{The difference of free energy density between the anisotropic state ($\omega \neq 0$) and the isotropic state ($\omega = 0$) for the fixed values of $z$ in (a) and $\alpha$ in (b).}
	\label{fig:dfreeED}
	\end{center}
\end{figure}

To check the behaviour of the free energy difference for the given testing ranges of $\alpha$ and $z$, we plot the free energy density difference 
by employing the following parameterizaion because the free energy itself shows very large magnitude: 
\begin{eqnarray}
{\rm T}(dF/V_{3})&\equiv& \tanh[\{\textrm{free energy density}(\omega \neq 0) - \textrm{free energy density}(\omega = 0) \}] \\ \nonumber
&=&\tanh[(F_{YM}  - F_{YM-iso} )/V_{3}].
\end{eqnarray}
The $(dF/V_{3})$ is negative means that $F_{YM} < F_{YM-iso} $ and so the anisotropic phase is stable. 

{In Fig.\ref{fig:dfreeED}-(a), a graph of $\rm T(df/V_3)$ versus $\alpha$(in the range of $0\leq \alpha \leq 5$) is given to show the free energy density difference between the anisotropic and the isotropic phase for the given values of $z$ in our testing range ($1 \leq z \leq 5$).  
$\rm T(df/V_3)$ is negative from $\alpha=0$ to around $\alpha = 3$ for the integral values of $z=1,2,3,4$ and $5$, but it turns to be positive after $\alpha \sim 3$. This change occurs abruptly around $\alpha = 3 \sim 4$, and $\rm T(df/V_3)$ shows large oscillation after that. Therefore, it is hard to distinguish the regions of anisotropic phase from the isotropic ones after this.
We conclude that for the given values of $z=1,2,3,4$ and $5$ and for the range of $0\leq\alpha\leq{\rm \ (about)\ }3$, the values of the chemical potential, $\mu_c$ that we obtain in Fig.\ref{zalphafix}-(a) are the critical values. 
After this region, however, all the curves switch their signs and it is clear that anisotropy is not the favoured states for any values of the chemical potential. When $\alpha \geq (\rm \ about\ )4$, there will not be anisotropy.

In Fig.\ref{fig:dfreeED}-(b), 
a $\rm T(df/V_3)$ versus $z$ graph(in the range of $1\leq z \leq 5$) is displayed for the given integral values of $\alpha=0,1,2,3$ and $4$.
For $\alpha=0,1,2$, $\rm T(df/V_3)$ is negative in such a range of $z$. The chemical potential addressed in Fig.\ref{zalphafix}-(b) is the critical value in $1 \leq z \leq 5$ for these values of $\alpha$. However, for $\alpha=3,4$, the free energy density difference, $dF$, is negative up to a certain value of $\alpha$ and then abruptly jumps to be positive. The turnning point of $\alpha$ is placed at $3 \leq z \leq 4$ for $\alpha=3$, and at $z \sim 1$ for $\alpha=4$. Thus, the anisotropic phase is favoured in $1\leq z \leq (\rm \ about\ )3$ for the chemical potential that we obtain in Fig.\ref{zalphafix}-(b).
The anisotropic phase could be stable just around $z \sim 1$ for $\alpha=4$, but the most of range of $z>1$ the isotropic phase is stable. This result is consistent with one for Fig.\ref{fig:dfreeED}-(a), where $\rm T(df/V_3)$ is positive after $\alpha \sim 4$ regardless of the values of $z$.

A closer look of the free energy density for the given integral values of $z$ is in Fig.\ref{fig:freeEDz}. The free energy densities for the anisotropic and isotropic phase are plotted together in Fig.\ref{fig:freeEDz}-(a),(c),(e). The red solid and the blue dashed lines indicate the anisotropic and isotropic phase respectively. Their differences are shown in Fig.\ref{fig:freeEDz}-(b),(d),(f) more in detail. It is clear that their signs of $\rm T(df/V_3)$ change roughly between $\alpha=2.5$ and $\alpha=3.8$.

For the fixed integral values of $\alpha$, the free energy densities are depicted in Fig.\ref{fig:freeEDa}-(a),(c),(e), where the light red solid lines and the purple dashed lines are for the anisotropic and isotropic phase respectively. When $\alpha=0,1,2$, 
$\rm T(df/V_3)$ is negative in the entire region of our test range of $z$ and so the anisotropic phases are favoured with the corresponding critical values of the chemical potential. 
However, for $\alpha=3$ in Fig.7-(c), 
the free energy of the anisotropic phase is smaller than that of the isotropic phase when $1 \leq z \leq {\rm(roughly)}3$ and so the anisotropic phase is favoured in this region. When after this, the isotropic phase is favoured.}

\begin{figure}[!h]
	\begin{center}
	\begin{subfigure}[b]{0.45\textwidth}
		\includegraphics[scale=0.55]{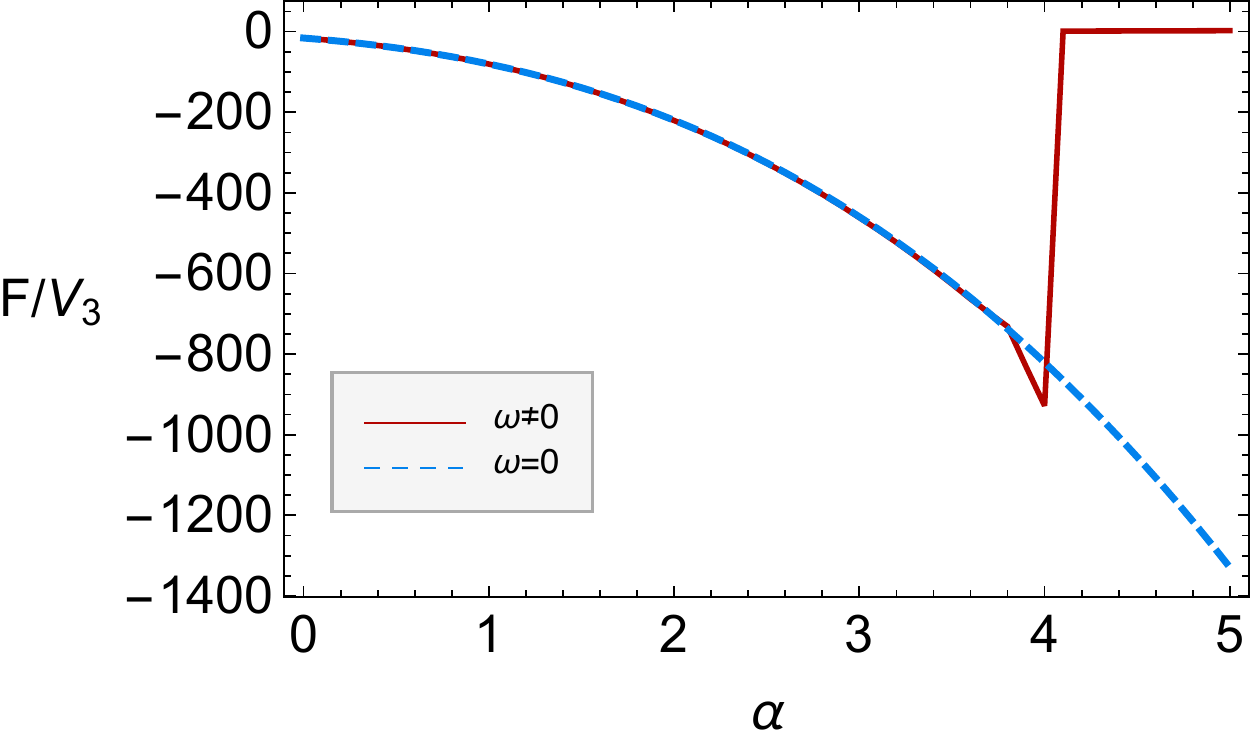}
		\caption{When $z=1$, the free energy density depending on $\alpha$}
	\end{subfigure} $\; \; \; $
	\begin{subfigure}[b]{0.45\textwidth}
		\includegraphics[scale=0.55]{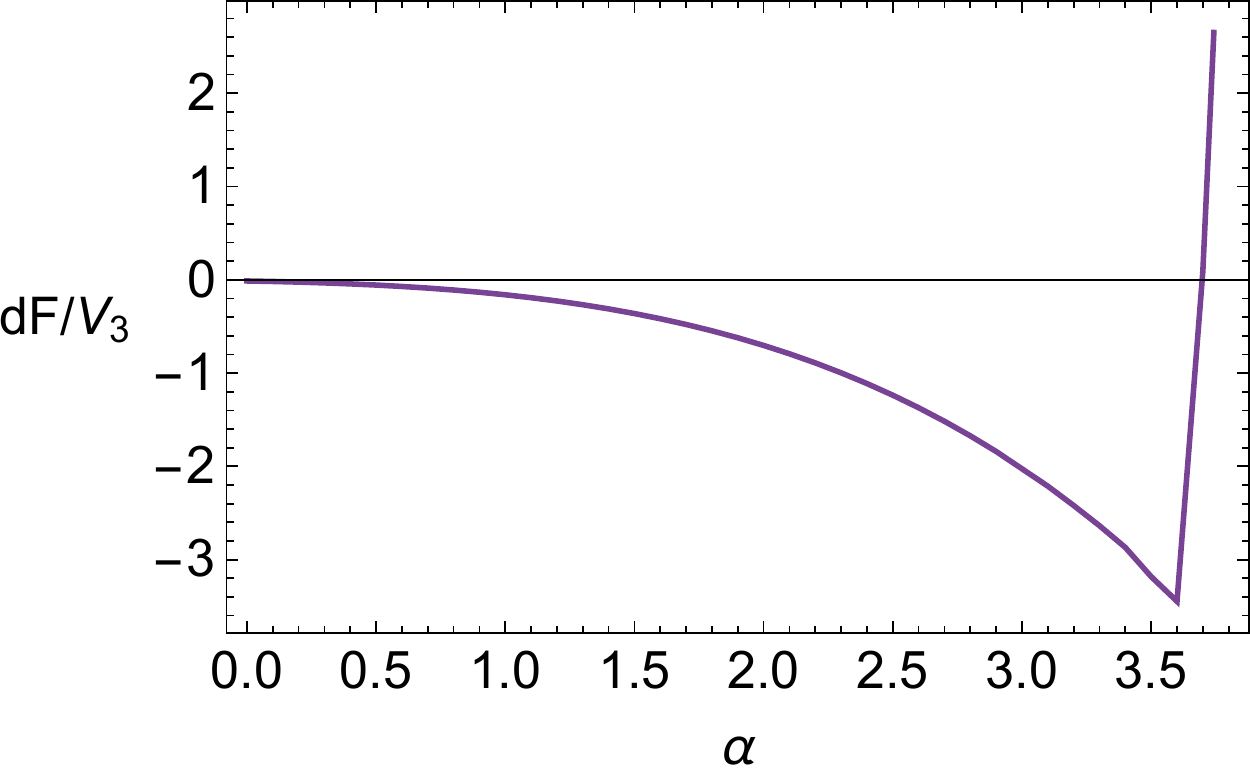}
		\caption{When $z=1$, the difference of the free energy density depending on $\alpha$}
	\end{subfigure}
	$\;$\\
	\begin{subfigure}[b]{0.45\textwidth}
		\includegraphics[scale=0.55]{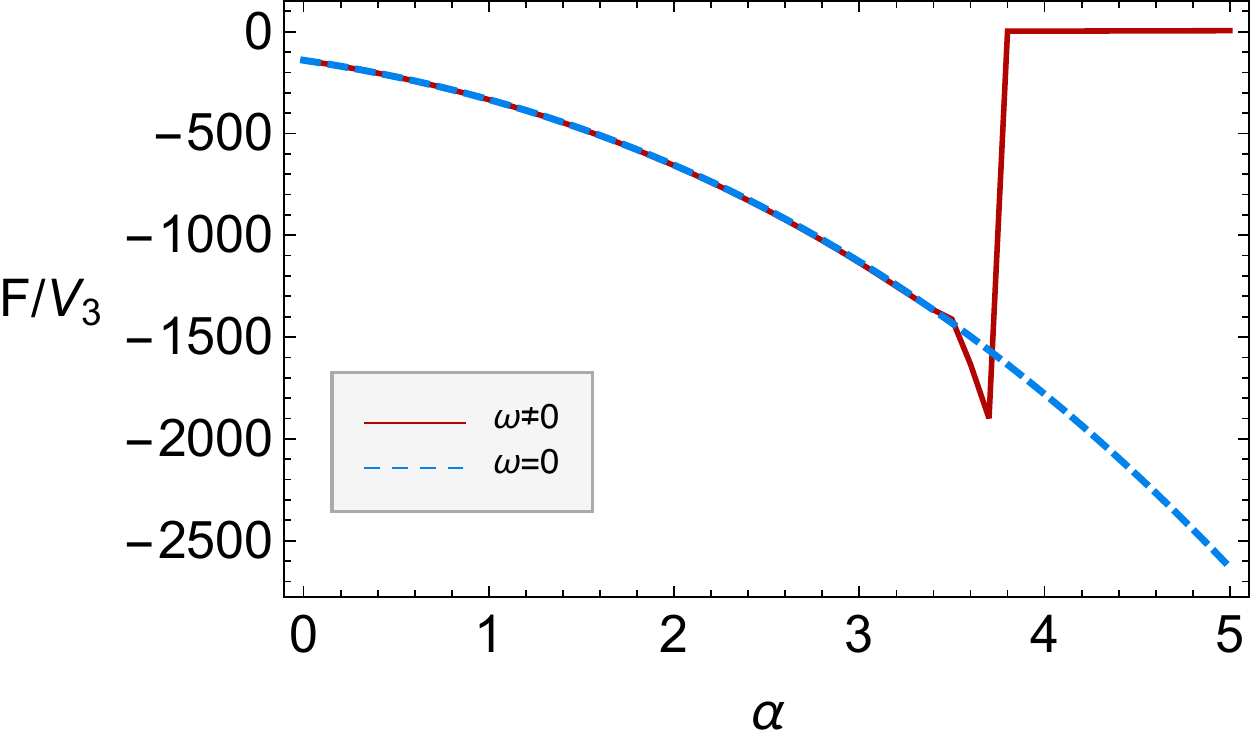}
		\caption{When $z=2$, the free energy density depending on $\alpha$}
	\end{subfigure}$\; \; \; $
	\begin{subfigure}[b]{0.45\textwidth}
		\includegraphics[scale=0.55]{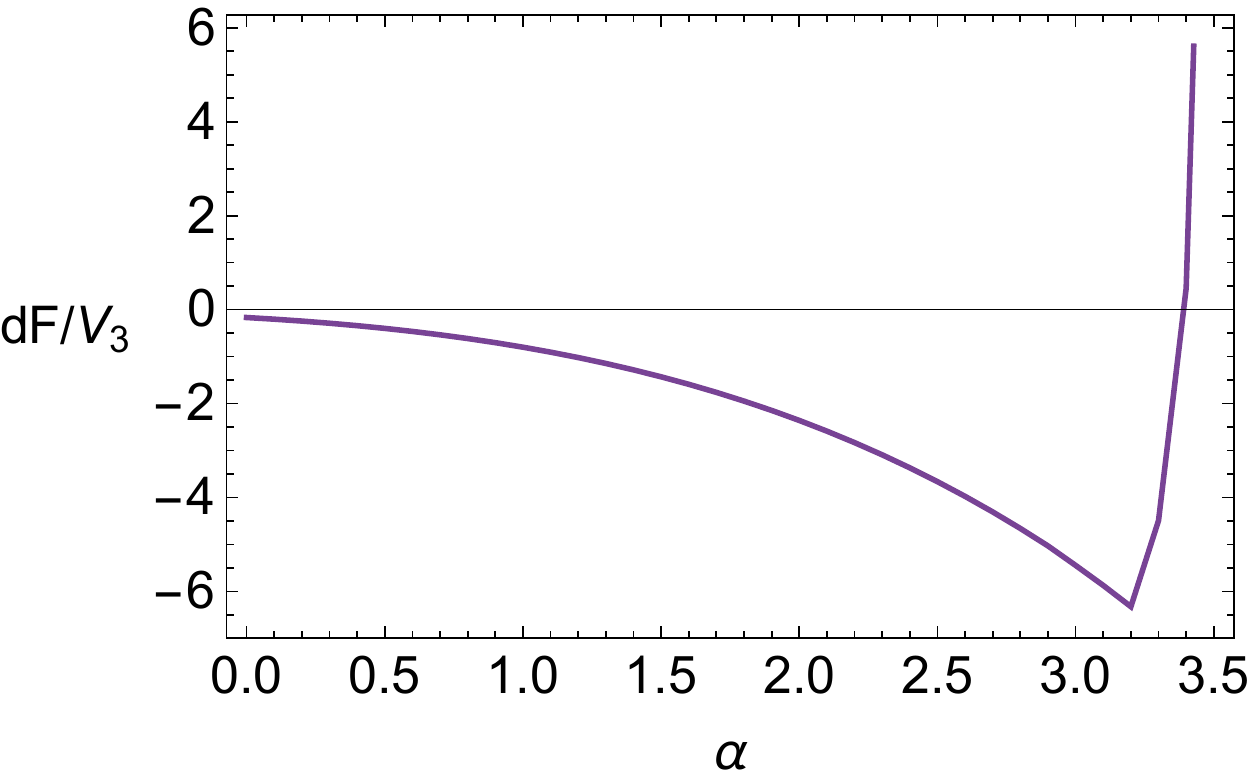}
		\caption{When $z=2$, the difference of the free energy density depending on $\alpha$}
	\end{subfigure}
	$\;$\\
	\begin{subfigure}[b]{0.45\textwidth}
		\includegraphics[scale=0.55]{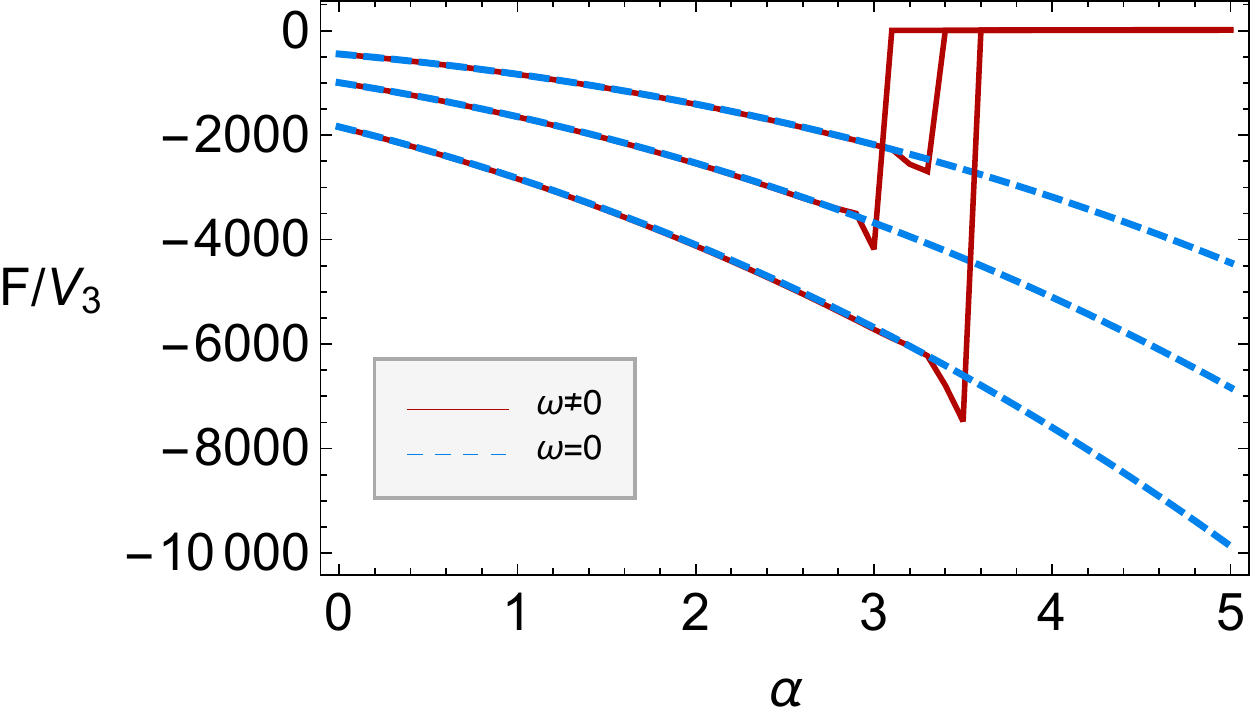}
		\caption{When $z=3,4,5$ from top to bottom on the left axis. The free energy density depending on $\alpha$ \\ $\;$ }
	\end{subfigure} $\; \; \; $
	\begin{subfigure}[b]{0.45\textwidth}
		\includegraphics[scale=0.55]{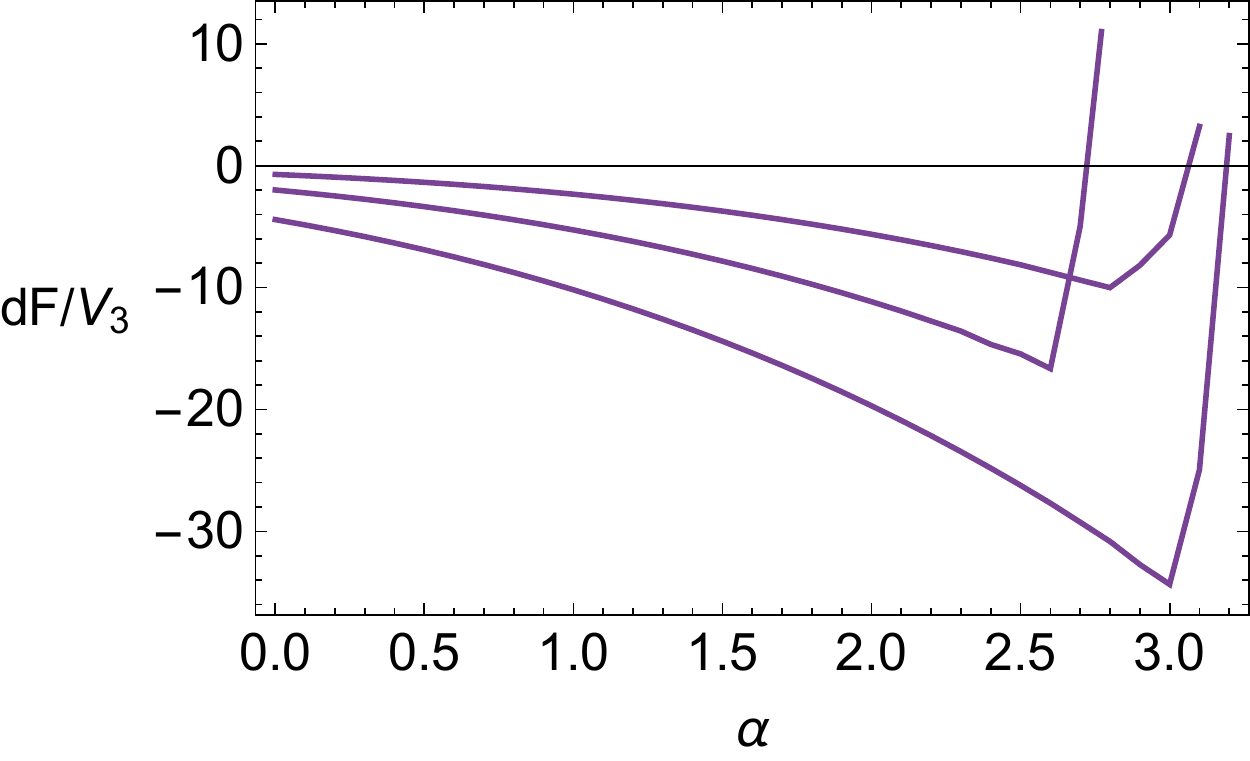}
		\caption{When $z=3,4,5$ from top to bottom on the left axis, the difference of the free energy density depending on $\alpha$}
	\end{subfigure}
	\caption{For the fixed value of $z$. The free energy densities depending on $\alpha$ for the anisotropic and isotropic phase(left), and the free energy density difference between the two phases(right).}
	\label{fig:freeEDz}
	\end{center}
\end{figure}

\begin{figure}[h!]
	\begin{center}
		\begin{subfigure}[b]{0.45\textwidth}
		\includegraphics[scale=0.6]{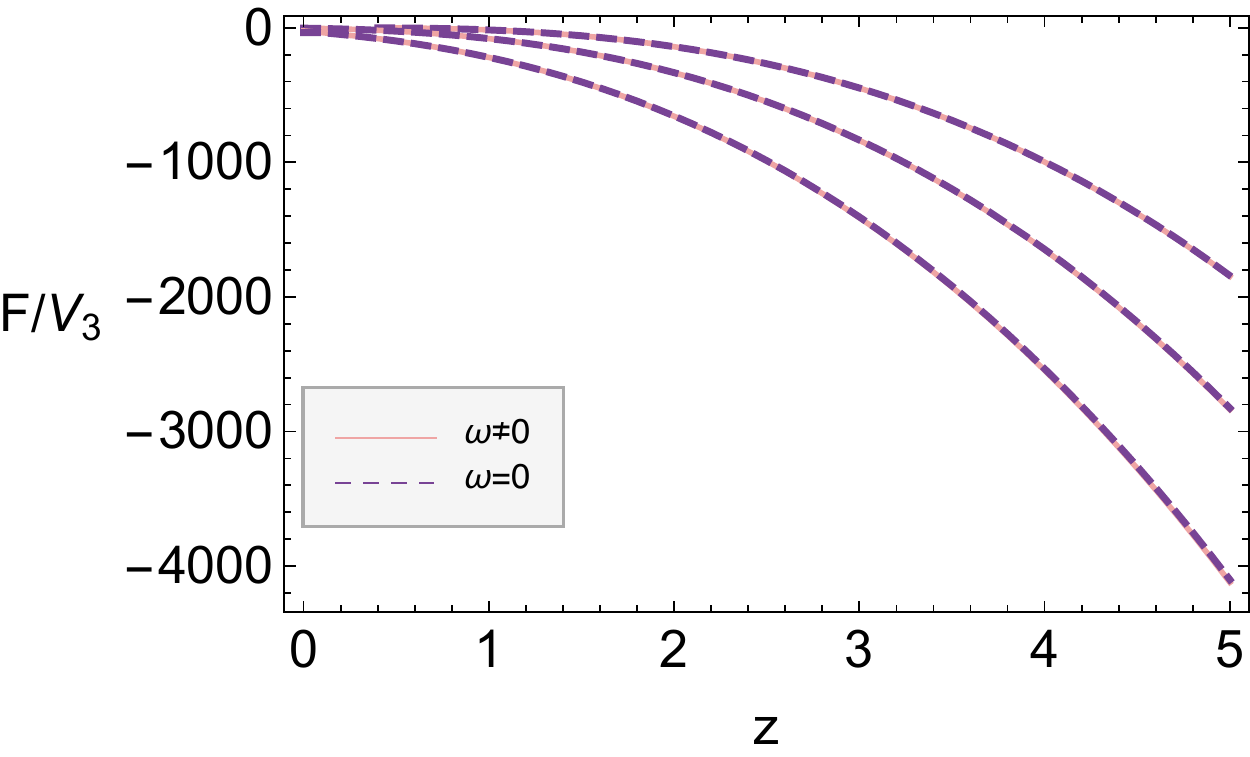}
		\caption{When $\alpha=0,1,2$ (from top to bottom on the right axis), the free energy density depending on $z$ }
	\end{subfigure}$\; \;$
	\begin{subfigure}[b]{0.45\textwidth}
		\includegraphics[scale=0.6]{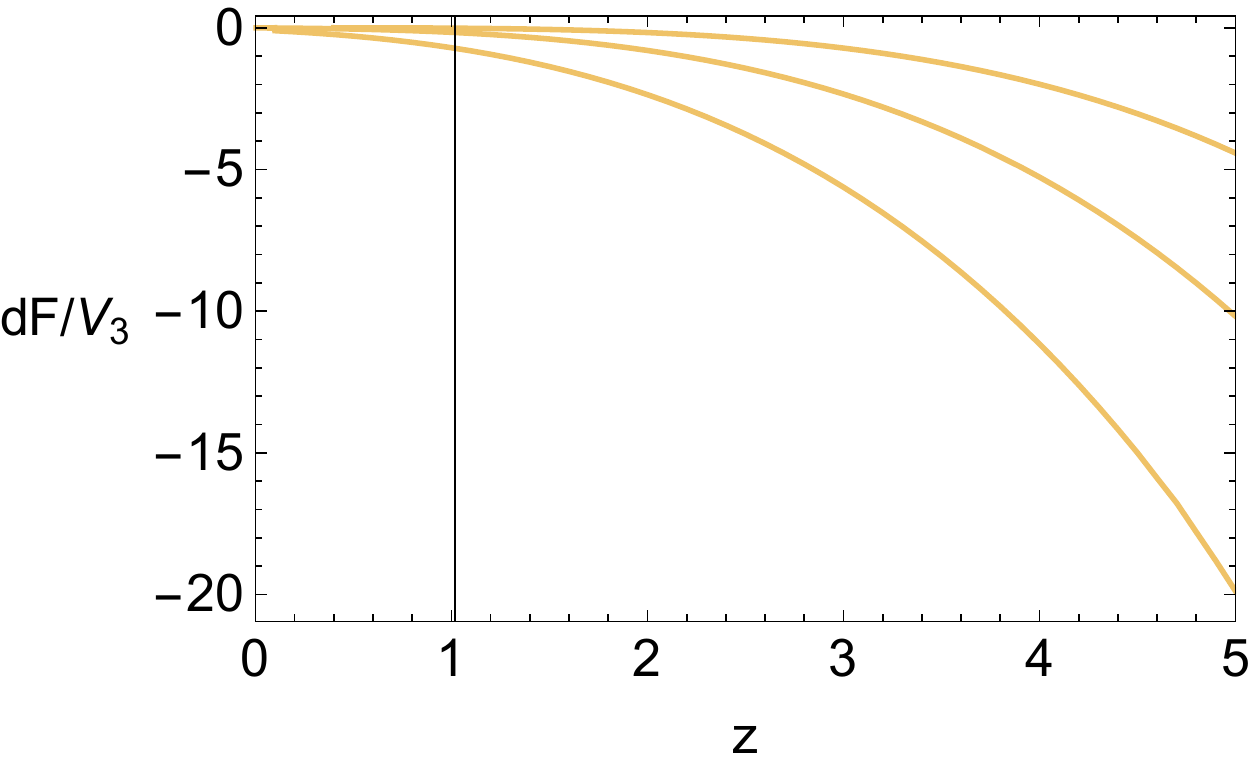}
		\caption{When $\alpha=0,1,2$ (from top to bottom on the right axis), the difference of the free energy density depending on $z$}
	\end{subfigure}
	$\;$\\
	\begin{subfigure}[b]{0.45\textwidth}
		\includegraphics[scale=0.6]{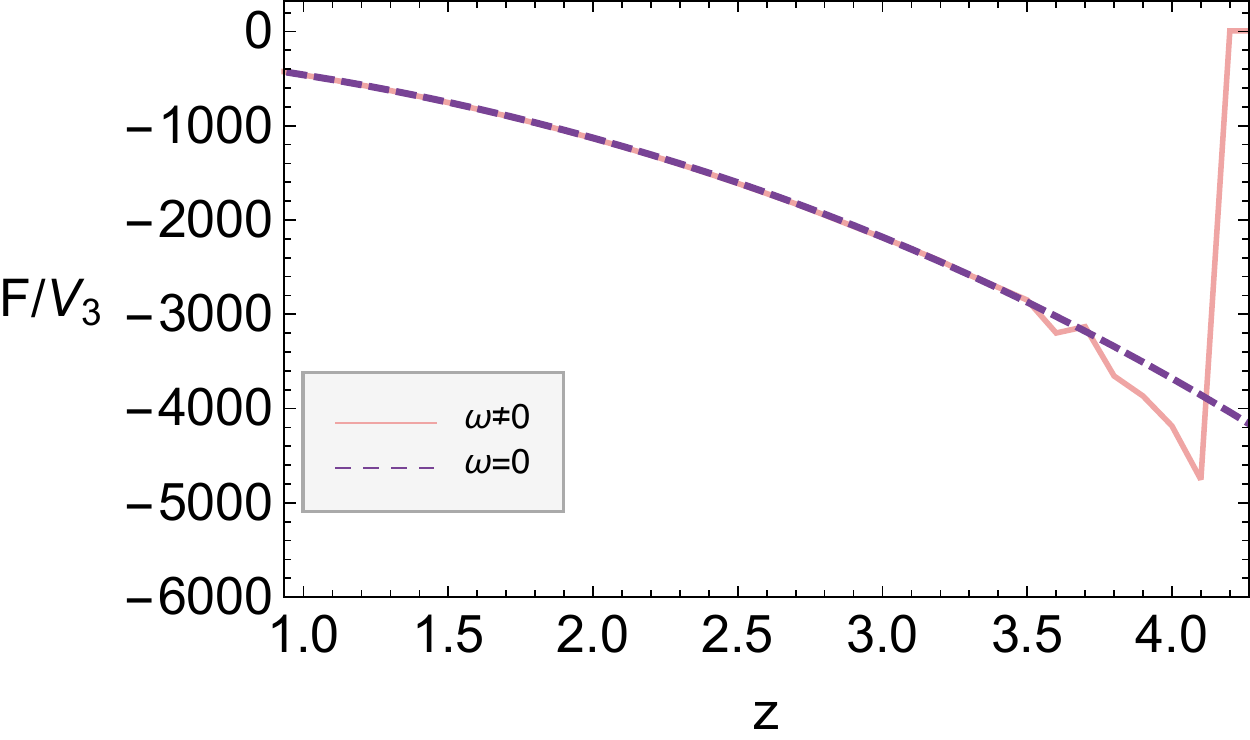}
		\caption{When $\alpha=3$, the free energy density depending on $z$}
	\end{subfigure}$\; \;$
	\begin{subfigure}[b]{0.45\textwidth}
		\includegraphics[scale=0.6]{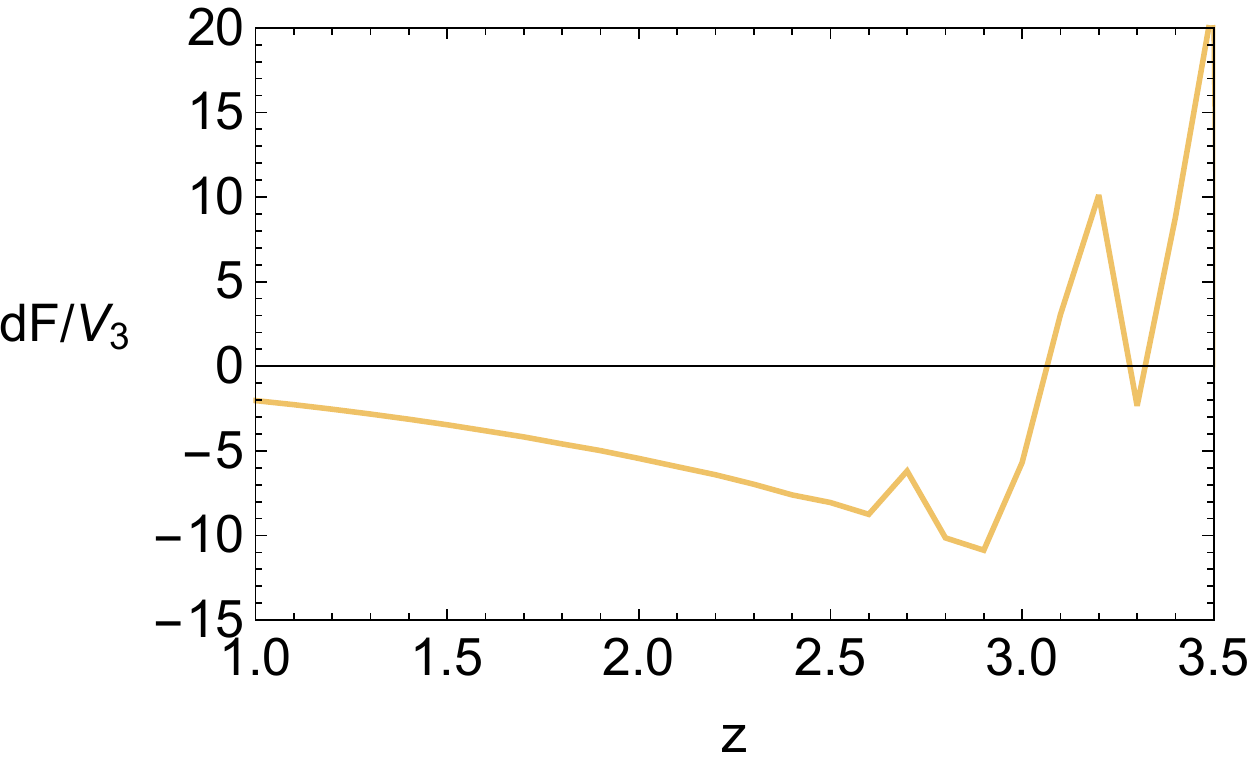}
		\caption{When $\alpha=3$, the difference of the free energy density depending on $z$}
	\end{subfigure}
	$\;$\\
	\begin{subfigure}[b]{0.45\textwidth}
		\includegraphics[scale=0.6]{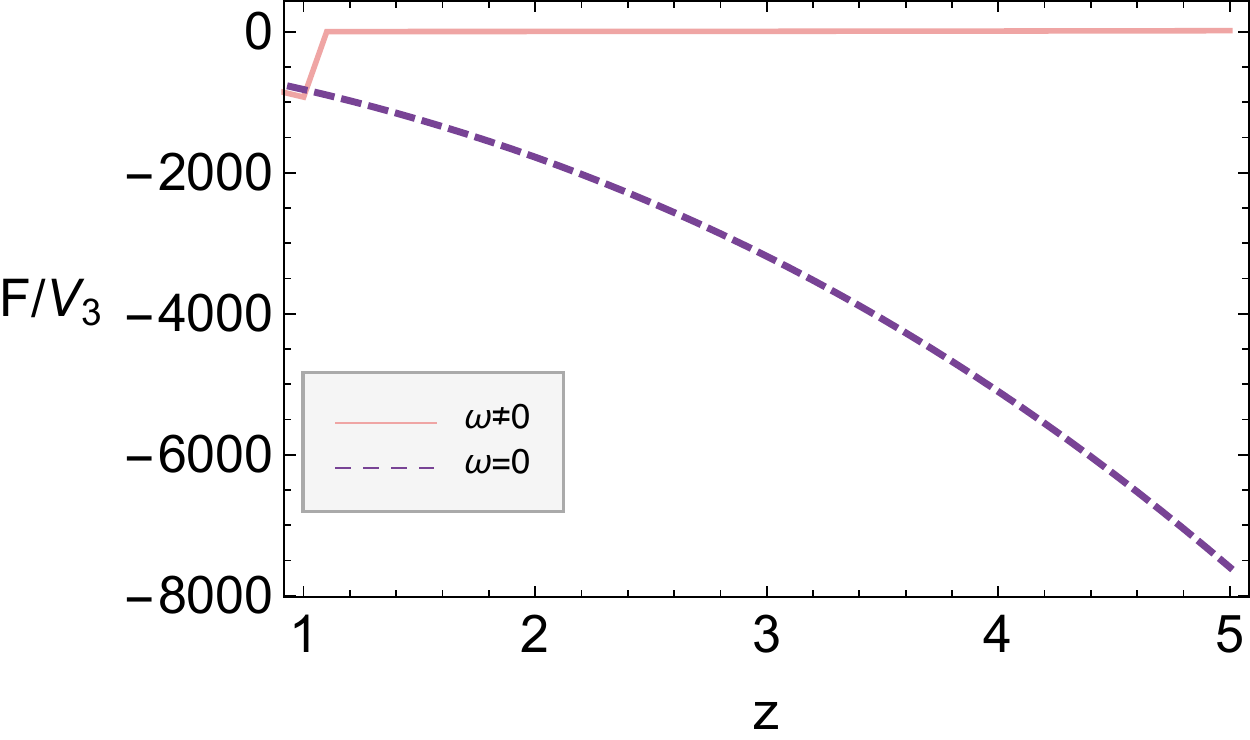}
		\caption{When $\alpha=4$, the free energy density depending on $z$}
	\end{subfigure}$\; \;$
	\begin{subfigure}[b]{0.45\textwidth}
		\includegraphics[scale=0.6]{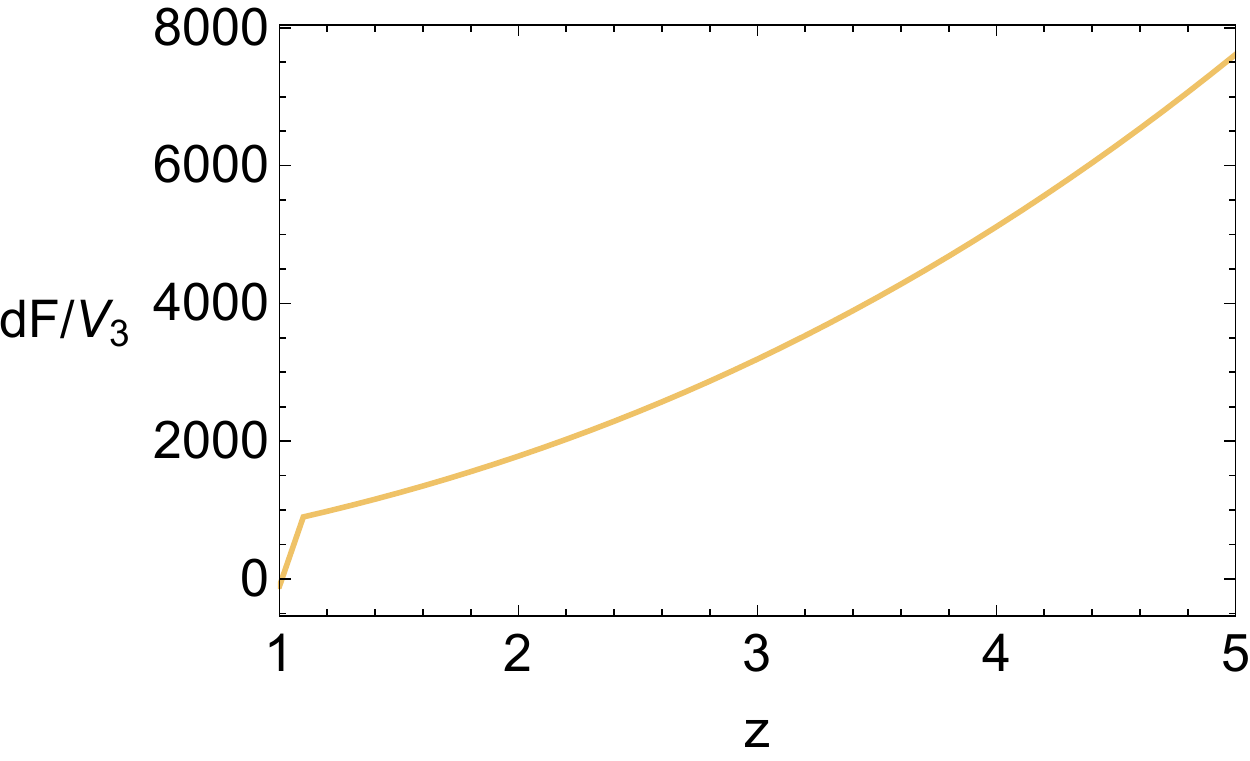}
		\caption{When $\alpha=4$, the difference of the free energy density depending on $z$}
	\end{subfigure}
	\caption{For the fixed value of $\alpha$. The free energy densities depending on $z$ for the anisotropic and isotropic phase(left), and the free energy density difference between two phases(right).}
	\label{fig:freeEDa}
	\end{center}
\end{figure}

\section{Summary}
\label{summary}

We explored thermodynamic phase transition between spatially isotropic and anisotropic phases of fluid dynamics by employing its gravity duals characterized by the hyperscaling violation factor $\alpha$ and the dynamical critical exponent $z$ when the current $\langle J^{x_{1}}_{1} \rangle$ starts to occur. We establish analytic and numerical methods to find the critical value of $\mu$ 
for generic values of $\alpha$ and $z$, and check the thermodynamic stability of the anisotropic phase by calculating the free energy. 

To do so, we employ its dual gravity action with the Einstein-dilaton-U(2) gauge fields and consider the probe limit that the Yang-Mills coupling constant is large, $\frac{\kappa^2_5}{g^2_{YM}}\rightarrow0$. We calculate the upper bounds of the critical value of the $\mu$ for the generic values of the $\alpha$ and $z$ 
by using the Sturm-Liouville method by using the test function (\ref{testftn}). The result is displayed in Fig.\ref{zalphafix} with the solid lines. We also calculate the critical value of $\mu$ by solving the coupled Yang-Mills field equations numerically. With a choice of the magnitude of the vector order parameter as $\langle J^{x_{1}}_{1} \rangle \sim \epsilon = 10^{-5}$, the shooting method searches the critical values of $\mu$ for the given values of $\alpha$ and $z$ satisfying appropriate boundary conditions. This result is shown in Fig.\ref{zalphafix} with the dashed lines. The two methods coincide within a few percetile errors as illustrated in Fig.\ref{fig:dffmu}. 

Next, we compute the free energy to check the thermodynamically favored phases among the spatially isotropic($\omega =0$) and the anisotropic($\omega \neq 0$) phases. {We investigate the region with $1 \leq z \leq 5$ and $0 \leq \alpha \leq 4$ and we found that} the anisotropic phase is stable only for {$0 \leq \alpha \leq (\rm roughly)3$ for all values of $z$, and the isotropic phase is favoured in the rest of the range of $\alpha$ and $z$. This result is shown in Fig.\ref{fig:dfreeED}, Fig.\ref{fig:freeEDz}, and Fig.\ref{fig:freeEDa}.} 

For our future work, it would be interesting to study the thermodynamic stability when considering the back reaction of Yang-Mills fields to the spacetime geometry, the dialton and the gauge fields to study this system beyond the probe limit. Furthermore, the back reation to the background metric will provide computation of the shear viscosity and its holographic renormalization. As argued in \cite{Oh:2012zu}, the shear viscosity of the anisotropic fluids runs as energy scale changes whereas almost of the other holographic models for fluid dynamics give the trivial flow of the shear viscosity. Exploring this for the generic values of $z$ and $\alpha$ would be interesting.

Following the study in \cite{Hollands:2012sf}, the positivity of canonical energy is equivalent to the dynamical instability and the canonical energy has the connection to the thermodynamic instability as well. Thus the checking the dynamical instability such as the quasinormal modes of this gravitational system with or without back reaction would be also interesting.

\section*{Acknowledgement}

M. Park is supported by TJ Park Science Fellowship of POSCO TJ Park Foundation. Jiwon Park is supported by Kwanjeong Fellowship of Kwanjeong Educational Foundation.
J.H.O thanks to his W.J. This research was supported by Basic Science Research Program through the National Research Foundation of Korea(NRF) funded by the Ministry of  Science, ICT $\&$ Future Planning(No.201600000001318).


\end{document}